\documentclass[12pt]{article}
\usepackage{epsfig}
\usepackage{comment}
\usepackage{latexsym}
\usepackage{color}
\newcommand{\mysquare}[0]{\raise-.2ex\hbox{{\Large$\Box$}}}
\def\lsim{\mathrel{\rlap {\raise.5ex\hbox{$ < $}}
{\lower.5ex\hbox{$\sim$}}}}
\def\gsim{\mathrel{\rlap {\raise.5ex\hbox{$ > $}}
{\lower.5ex\hbox{$\sim$}}}} \topmargin -1.5cm \textheight=22.5cm \textwidth=16.5cm
\catcode`\@=11
\newcount\hour
\newcount\minute
\newtoks\amorpm
\hour=\time\divide\hour by60 \minute=\time{\multiply\hour by60 \global\advance\minute by-\hour}
\edef\standardtime{{\ifnum\hour<12 \global\amorpm={am}%
        \else\global\amorpm={pm}\advance\hour by-12 \fi
        \ifnum\hour=0 \hour=12 \fi
        \number\hour:\ifnum\minute<10 0\fi\number\minute\the\amorpm}}
\edef\militarytime{\number\hour:\ifnum\minute<10 0\fi\number\minute}
\def\draftlabel#1{{\@bsphack\if@filesw {\let\thepage\relax
   \xdef\@gtempa{\write\@auxout{\string
      \newlabel{#1}{{\@currentlabel}{\thepage}}}}}\@gtempa
   \if@nobreak \ifvmode\nobreak\fi\fi\fi\@esphack}
        \gdef\@eqnlabel{#1}}
\def\@eqnlabel{}
\def\@vacuum{}
\def\draftmarginnote#1{\marginpar{\raggedright\scriptsize\tt#1}}
\def\draft{\oddsidemargin -.2truein
        \def\@oddfoot{\sl preliminary draft \hfil
        \rm\thepage\hfil\sl\today\quad\militarytime}
        \let\@evenfoot\@oddfoot \overfullrule 3pt
        \let\label=\draftlabel
        \let\marginnote=\draftmarginnote
   \def\@eqnnum{(\theequation)\rlap{\k

 ern\marginparsep\tt\@eqnlabel}%
\global\let\@eqnlabel\@vacuum}  }
\newcommand{\be}[0]{\begin{equation}}
\newcommand{\ee}[0]{\end{equation}}
\newcommand{\ba}[0]{\begin{eqnarray}}
\newcommand{\ea}[0]{\end{eqnarray}}

\def\bs{\begin{subequations}}
\def\es{\end{subequations}}

\def\thebibliography#1{%
\vskip 0.5cm \centerline{\bf \Large References}
\list{%
[\arabic{enumi}]}{\settowidth\labelwidth{[#1]} \leftmargin\labelwidth \advance\leftmargin\labelsep
\usecounter{enumi}}
\def\newblock{\hskip .11em plus .33em minus .07em}
\sloppy\clubpenalty4000\widowpenalty4000 \sfcode`\.=1000\relax}

\renewcommand{\theequation}{\arabic{section}.\arabic{equation}}

\renewcommand{\section}{\setcounter{equation}{0}\@startsection
{section}{1}{0mm}{-\baselineskip}{0.5\baselineskip} {\normalfont\Large\bfseries}}

\renewcommand{\subsection}{\@startsection
{subsection}{2}{0mm}{-\baselineskip}{0.5\baselineskip} {\normalfont\large\bfseries}}

\renewcommand{\subsubsection}{\@startsection
{subsubsection}{3}{0mm}{-\baselineskip}{0.5\baselineskip} {\normalfont\normalsize\slshape}}

\usepackage{amssymb,amsfonts}
\usepackage{graphicx}
\usepackage{cite}

\newcommand{\ie}{{\em i.e. }}

\newcommand{\Z}{\mathbb{Z}}

\renewcommand{\O}{{\cal O}}
\renewcommand{\Re}{{\rm Re}}
\renewcommand{\Im}{{\rm Im}}

\newcommand{\with}{\mbox{with}}
\newcommand{\when}{\mbox{when}}
\renewcommand{\and}{\mbox{and}}

\newcommand{\tg}{\tilde{g}}
\newcommand{\tm}{\tilde{m}}
\newcommand{\F}{{\cal F}}
\newcommand{\T}{{\cal T}}
\newcommand{\U}{{\cal U}}
\newcommand{\V}{{\cal V}}

\renewcommand{\L}{{\cal L}}
\newcommand{\bQ}{{\bar Q}}
\newcommand{\bL}{{\bar L}}
\newcommand{\bepsilon}{{\bar \epsilon}}
\newcommand{\Th}[2]{\theta\left[^{#1}_{#2}\right]}

\begin{document}
%\verb|\usepackage{draftcopy}|\\
\begin{titlepage}
\begin{flushright}
LPTENS--08/42, 
CPHT--RR054.0708, January 2009
\end{flushright}

%\vspace{2mm}

\begin{centering}
{\bf\huge Induced superstring cosmologies}\\
\vspace{2mm}
{\bf\huge and moduli stabilization$^\ast$}\\

\vspace{4mm}
 {\Large Tristan Catelin-Jullien$^{1}$, Costas~Kounnas$^{1}$\\
 Herv\'e~Partouche$^2$ and Nicolaos~Toumbas$^3$}

\vspace{2mm}

$^1$ Laboratoire de Physique Th\'eorique,
Ecole Normale Sup\'erieure,$^\dagger$ \\
24 rue Lhomond, F--75231 Paris cedex 05, France\\
{\em  catelin@lpt.ens.fr, Costas.Kounnas@lpt.ens.fr}

\vskip .1cm

$^2$ {Centre de Physique Th\'eorique, Ecole Polytechnique,$^\diamond$
\\
F--91128 Palaiseau, France\\
{\em Herve.Partouche@cpht.polytechnique.fr}} \vskip .1cm

$^3$ {Department of Physics, University of Cyprus,\\ Nicosia 1678, Cyprus }\\ {\em nick@ucy.ac.cy}
 \vspace{1.5mm}

{\bf\Large Abstract}

\end{centering}
\vspace{2mm}
%\begin{quote}
\noindent 
We extend the analysis of the recently obtained stringy cosmological solutions induced by thermal and quantum effects, once space-time supersymmetry is spontaneously broken by geometrical fluxes. Cases in which more than one modulus participating in the supersymmetry breaking mechanism are investigated. The free energy is obtained at the full string level. In the intermediate cosmological region where the temperature and the supersymmetry breaking scale are sufficiently smaller than the Hagedorn temperature, the quantum and thermal corrections are under control and calculable.
The reason is that the contributions to the effective potential of the moduli that are not participating in the supersymmetry breaking are exponentially suppressed. 
The backreaction on the initially flat background results in many cases into cosmological evolutions, where the dynamics of all complex structure moduli is frozen. The solutions describe effectively a radiation dominated era, where thermal effects are never negligible, even if the temperature tends to zero at late times. We analyze several types of supersymmetry breaking patterns and examine the stability of the corresponding radiation era. 
%\end{quote}

\vspace{5pt} \vfill \hrule width 6.7cm \vskip.1mm{\small \small \small \noindent $^\ast$\ Research
partially supported by the EU (under the contracts MRTN-CT-2004-005104, MRTN-CT-2004-512194,
MRTN-CT-2004-503369, MEXT-CT-2003-509661), ANR (CNRS-USAR) contract 05-BLAN-0079-02,
 CNRS PICS 2530, 3059 and 3747, INTAS grant 03-51-6346 and  INTERREG IIIA Crete/Cyprus.\\
$^\dagger$\ Unit{\'e} mixte  du CNRS et de l'Ecole Normale Sup{\'e}rieure associ\'ee \`a
l'Universit\'e Pierre et Marie Curie (Paris
6), UMR 8549.\\
 $^\diamond$\ Unit{\'e} mixte du CNRS et de l'Ecole Polytechnique,
UMR 7644.}

\end{titlepage}
\newpage
\setcounter{footnote}{0}
\renewcommand{\thefootnote}{\arabic{footnote}}
 \setlength{\baselineskip}{.7cm} \setlength{\parskip}{.2cm}

\setcounter{section}{0}

%%%%%%%%%%%%%%%%%%%%%%%%%%%%%%%%%%%%
%%%%%%%%%%%%%%%%%%%%%%%%%%%%%%%%%%%%

\section{Introduction}

 \noindent
Perhaps the most natural setting for testing string theory is the cosmology 
of our Universe. By now, there is a plethora of cosmological data favoring the 
phenomenological model of hot Big Bang cosmology, where the cosmological 
evolution starts with a highly singular event, the Big Bang, followed by an initial period of rapid inflation, 
a very high temperature phase, a proportionally large amount of dark matter and dark energy \cite{Perlmutter}.  
Many features of this phenomenological model are not well understood,  
and the hope is that their explanation will arise from a fundamental 
theory of quantum gravity such as string theory.  

\noindent
In an effort to build a concrete theoretical framework for studying cosmology,
a class of string theory vacua, where the backreaction of both thermal and quantum
effects can be systematically taken into account, was recently examined in \cite{ktt, kp1, kp2, ckpt}. 
In particular, starting with  weakly coupled four-dimensional heterotic string models, with initial
$N=4$ or $N=2$ space-time supersymmetry \cite{ckpt}, and implementing
the thermal and the quantum corrections due to the spontaneous breaking of supersymmetry,
cosmological solutions are found, at least when the temperature $T$ and the supersymmetry
breaking scale $M$ are sufficiently below the Hagedorn temperature $T_H$. The string coupling 
must also be sufficiently small so that perturbative, thermal equilibrium computations of string theory
can be applied.   
In these examples, all moduli that are not involved in the breaking of supersymmetry give
exponentially suppressed contributions to the (thermal) effective potential. 
This is essentially the underlying reason for the no-scale structure \cite{kp2, ckpt} characterizing 
the models. 
Due to this remarkable property,  
the thermal and quantum corrections are under control and calculable. 
We would like to stress here one of the merits of the one-loop string computations we perform: 
namely, the absence of both infrared and ultraviolet ambiguities 
leads to a well defined energy density and pressure \cite{ckpt}.    

\noindent
The backreaction of the thermal and quantum corrections on the initially flat metric 
and on certain moduli fields (including the dilaton field and the geometrical moduli participating 
in the supersymmetry breaking mechanism)
induces the cosmological evolution. Both the temperature and the supersymmetry breaking scale
evolve in time inversely proportional to the scale factor of the Universe \cite{ckpt}. 
As the Universe expands, the system remains weakly coupled in the future, 
and so we can trust the solution at later times. 
However, if we extrapolate the solution
back in time, the temperature rises and as this approaches the Hagedorn
temperature, we encounter the Hagedorn instabilities of
thermal string theory, indicating
a non-trivial high temperature phase transition \cite{AW,RK,AKADK}. At this point
the dynamics drive the system outside the perturbative domain, and the large backreaction of the
tachyon condensates involved in the transition may drive the system to strong coupling. 
In the literature there are
some speculative proposals concerning the nature of this early times/high temperature phase 
transition \cite{AW,RK,AKADK,BV,CosmoTopologyChange,massivesusy,akpt}, 
but still an adequate quantitative description of the dynamics is lacking. We will assume that such a
stringy phase transition occurs at very early times, and that the system eventually exists in a low temperature, 
$T<<T_H$,
weakly coupled phase, with three large spatial directions and possibly some intermediate size spatial directions that 
generate the scale of supersymmetry breaking, $M<<T_H$, via geometrical fluxes. 
In this low temperature phase, we have good analytical control to analyze the subsequent 
cosmological evolution, using string perturbation theory techniques.  

\noindent
To be more precise, we are forced to separate the cosmological evolution in four distinct phases, according to the value of the temperature, namely:

\noindent
$(i)$ The very early
``Big Bang" phase, where the underlying string theory degrees of freedom are strongly coupled. Perhaps string dualities can be applied to understand this phase and resolve the classical Big Bang singularity. 
See e.g. \cite{GV,CosmoTopologyChange,Khoury,Liu,CSV,massivesusy,akpt} and references therein. 

\noindent
$(ii)$ A high temperature stringy phase, $T \lesssim T_H$, where string oscillators and winding states must be properly taken into account \cite{AW,RK,AKADK,BV,Easther}. Often, these lead to a non-geometrical structure of the Universe, e.g. the T-fold cosmologies 
of \cite{ktt}, or even to a change of  the topology  and dimensionality of space \cite{CosmoTopologyChange,massivesusy}.

\noindent
$(iii)$ The third phase has features similar to that of a standard, radiation dominated Friedmann cosmology. Here, the Universe has cooled down to temperatures far below Hagedorn.  The effects of string massive states are exponentially suppressed ${\cal O}(\exp({-M_s / T }))$. In this phase, the ratio of the temperature  $T$  and
supersymmetry breaking scale $M$ is fixed, both evolving inversely proportional to the scale
factor of the Universe \cite{kp2, ckpt}. In cases with ${ N}=1$ initial supersymmetry, the behavior can be that of a cosmological constant dominated inflationary universe \cite{kp2,kp1}. 

\noindent 
$(iv)$ At lower temperatures, new phenomena such as the electroweak phase transition, QCD confinement  and structure formation are taking place. We expect also that in this phase, some dynamics becoming relevant at these lower temperatures will stabilize the
no-scale modulus associated to the supersymmetry breaking scale \cite{Noscale}.

\noindent 
Some interesting ideas concerning the first two stringy phases have been presented recently in \cite{akpt}, where it was argued that the introduction of certain chemical potentials in the standard canonical ensemble of superstrings removes the Hagedorn instabilities. These ensembles are characterized by a ``Temperature duality,"
${\cal Z}(T/T_H)={\cal Z}(T_H/T)$. (See also \cite{ACI} for some related work.) Equally interesting are the models of \cite{massivesusy},  which possess a new kind of {\it massive boson-fermion degeneracy symmetry}. Type II, heterotic and orientifold models have been presented. Some proposals are put forward, in the framework of these theories, concerning the early structure of the Universe. 

\noindent
Here however, we would like to examine more thoroughly the generalization of supersymmetry breaking in the cases where more moduli are involved, not only in heterotic string but also in type II closed string theories, working in the intermediate region $(iii)$. It is interesting that this intermediate phase can be studied with high precision 
at the full string level\cite{ckpt}, thanks to to the fact that just below the Hagedorn temperature, the theory possesses a no-scale structure.
The free energy is set by a single, 
overall scale, which can be chosen to be either the temperature or the supersymmetry 
breaking scale, and the rest of the dependence is given in terms of functions of dimensionless, complex structure like ratios. The backreaction on the initially flat background induces the cosmological evolution \cite{kp2, ckpt}.

\noindent
In the string models studied in \cite{ckpt}, only a single modulus was participating 
in the supersymmetry breaking mechanism. In this work we extend the analysis for
cases where more geometrical moduli participate in the spontaneous breaking of supersymmetry. As we already remarked, all such moduli give non-exponentially suppressed contributions to the effective potential. We examine in more details the case where the supersymmetry breaking arises via geometrical fluxes \cite{GeoFluxes} induced by stringy 
Scherk-Schwarz \cite{SSstringy} boundary conditions along two internal spatial cycles with radii denoted by $R_4$ and $R_5$. 
When the complex structure ratio $\U \equiv R_4/R_5$ is stabilized to be of order one, 
the supersymmetry breaking scale $M$ will be proportional to the inverse of the volume modulus $\sqrt{R_4R_5}$.
(The whole $\U$ dependence of the one-loop effective potential is presented and analyzed in this work.)
The results can be easily extended to cases involving more internal cycles.
The heterotic or type II geometrical fluxes are introduced by utilizing the helicity and/or other internal R-symmetry charges. In the type II models, left-moving and right-moving R-symmetry charges can be coupled symmetrically or asymmetrically to the two cycles. 

\noindent
As we will see, the low energy dynamics of some models admits a solution describing a radiation dominated era. During the evolution, the supersymmetry breaking scale $M$ and the temperature $T$ evolve in time the same way as the inverse of the scale factor of the Universe. It is then important to analyze the stability for the dimensionless modulus describing the shape of the internal space, $\U \equiv R_5/R_4$. We find a rich structure of phenomena, depending on whether the corresponding effective potential admits a minimum, a maximum or a run-away behavior. Complementary results can be found in \cite{BKP}. There, it is shown that these solutions are attractors for the dynamics. That is, there are basins of initial conditions whose associated cosmological evolutions converge towards the radiation dominated era with stabilized complex structures. The results of \cite{BKP} are important for the following reason. Due to the lack of quantitative control, 
the ambiguities of the Hagedorn transition exit can be parameterized in terms of initial time boundary conditions, ``capping off'' the cosmology at an early time just after the temperature has dropped below Hagedorn. It is then important to know that for generic initial conditions, the subsequent cosmological evolution is described by a solution such as the one displayed in this paper. We expect that the degrees of freedom describing the physics of the early times/high temperature stringy Hagedorn phase, decouple at later times and low temperatures.
   
\noindent
In this work we choose to break supersymmetry via geometrical fluxes which can be implemented consistently not only 
in the effective no-scale supergravity theory, but also at the string one-loop level. 
Our work is concerned with the intermediate cosmological
region $(iii)$ just before the electroweak phase transition. In this regime, the geometrical moduli $R_n$ that do not participate in the
supersymmetry breaking mechanism (that is, moduli other than the dilaton and the radial moduli $R_4$ and $R_5$) can:\\
- Either get a soft breaking mass at tree level, see for instance \cite{SSstringy}. They are thus thermally fluctuating about their ground state.\\
- Or, they remain massless at tree level.\\
 Both types of such moduli 
give exponentially suppressed contributions to the one-loop effective potential when their vevs are of order unity as
compared to the string scale \ie $R_n \ll  1/T,~ 1/M$. In \cite{BEKP}, it is explicitly shown that if this condition is initially satisfied right after the exit from the Hagedorn phase, it remains so during the whole  intermediate radiation dominated era {\em (iii)}\footnote{To be precise, this is true as long as the initial time-derivative of $R_n$ does not exceed some ``escape velocity'', above which the modulus enters in a different regime where it is spontaneously decompactified. In the latter case, the system is driven into a radiation-like era in higher dimensions.}. If we turn on kinetic energies for these massless moduli, these give sub-dominant contributions to the effective Hubble equation, of the order $1/a^6$. 
The same is valid for Wilson line moduli. The values of Wilson lines associated to extended symmetry points are 
preferred at finite temperature since these points give rise to more massless degrees of freedom.
It is important however to analyze and provide a mechanism for the
stabilization of relevant moduli just after the electroweak phase transition. One of these is the no-scale
modulus that sets the supersymmetry breaking scale and therefore the electroweak symmetry breaking scale via
radiative corrections. In this work we do not carry out this later time analysis. We also expect that the use of other
types of fluxes for moduli stabilization (which at the moment cannot be fully implemented at the string perturbative one-loop level) 
will result in similar features in the intermediate cosmological region we are working with, as was suggested in \cite{kp2}. 

\noindent
The paper is organized as follows. In Section $2$, we  
describe the basic principles of the construction, clarifying the domain of validity of our analysis. We examine different classes heterotic and type II string vacua and 
implement various ways that lead to a spontaneous breaking of supersymmetry. 
We calculate the free energy at the full string level, and obtain 
the generic structure of the thermal effective potential in the intermediate region where $T \ll T_H$. We use these general results to find their counterparts at zero temperature. We also  derive in the Einstein frame  the gravitational equations and the equations of motion 
for the main moduli participating in the supersymmetry breaking mechanism, 
including the dilaton field. We present solutions at zero temperature, where the time evolution of the scale factor and the stability of the complex structure $\U$ are analyzed.

\noindent
In Section $3$, we present explicit thermal models leading to radiation dominated cosmological evolutions.  
The compatibility between the gravitational equations and the equations of motion of 
the relevant moduli leads to the equation of state
$\rho=(3+n)\, P$, where $n$ is the number of internal dimensions involved in
the supersymmetry breaking mechanism. 
In addition, we analyze the stability for the modulus $\U$.
 We compare to the zero temperature situation, mainly to show the relevance 
of the thermal corrections to the cosmological evolution. We show that during the radiation era, even when the temperature is very small, thermal effects are never negligible.  The qualitative behaviors of thermal and non-thermal evolutions are drastically different. 
This is due to the non-linear character of the gravitational and moduli equations.

\noindent
The final Section is devoted to conclusions and perspectives.

%%%%%%%%%%%%%%%%%%%%%%%%%%%%%%%%%%%%
%%%%%%%%%%%%%%%%%%%%%%%%%%%%%%%%%%%%

\section{General setup}
We consider initially supersymmetric and weakly coupled flat backgrounds within the context of
four-dimensional superstrings constructed via orbifolds\cite{orbifolds} and/or via the free fermionic construction\cite{ABK}. 
By turning on certain non-trivial geometrical fluxes, we can spontaneously break 
space-time supersymmetry \cite{SSstringy}.
The procedure that we follow involves
coupling some of the internal lattice quantum numbers to the space-time Fermion number
$F$ and/or to any of the discrete R-symmetry charges $Q_R$. This is a generalization of the Scherk-Schwarz mechanism to superstrings\cite{SSstringy}.   
In addition, the system is put at finite temperature\cite{AW, RK, AKADK}. 
The temperature and the supersymmetry breaking scales are taken to be below the Hagedorn temperature.
Our aim is to study cosmological evolutions induced by
the thermal and quantum corrections, as in \cite{ckpt}, and in particular investigate 
how some of the geometrical moduli, participating in the breaking of supersymmetry, 
can be stabilized around a local minimum.
  
\subsection{Heterotic models}
We start with heterotic string compactifications on six-manifolds of the form 
$S^1(R_4)\times S^1(R_5)\times
{\cal M}_4$. Here, the choice of the four-manifold ${\cal M}_4$ 
determines the initial amount of supersymmetry: $N_4=4$ for the case of $T^4$, and 
$N_4=2$ for the cases of $T^4/\Z_2$ orbifold and $K_3$ compactifications.
Two internal cycles, associated to the $X^4$ and $X^5$ directions, have been singled out, since these are to be utilized to break supersymmetry spontaneously. 

\noindent
We illustrate the derivation of the pressure with the simplest example. The initially supersymmetric string partition function is given by
\be
 Z = \frac{V_3}{(2\pi)^3}~\int_\F {d^2\tau\over
2\tau_2^{5/2}}\ \Gamma_{(1,1)}(R_0) 
~\frac 12 \sum_{a,b} (-)^{a+b+ab} \Th ab^4~{\Gamma_{(6,22)} \over \eta^{12}\bar\eta^{24}}~,
\ee
where the Euclidean time direction is compactified on a circle of radius $R_0$. $V_3$ is the volume of the three large spatial directions.
The $\Gamma_{(6,22)}$ lattice is associated to the zero mode contribution of the internal six-manifold, along with the $E_8\times E_8$ or $SO(32)$ right-moving lattice. For instance, the $E_8\times E_8$ case on $T^6$ gives the block
\be
\Gamma_{(6,22)} = \Gamma_{(6,6)}~{1\over
2}\sum_{\gamma,\delta}\bar\theta[^\gamma_\delta]^8~{1\over
2}\sum_{\gamma',\delta'}\bar\theta[^{\gamma'}_{\delta'}]^8\, ,
\ee
where $\gamma, ~\delta$ and $\gamma',~ \delta'$ are integers defined modulo $2$.

\noindent
We wish to implement a non-zero temperature in the model. This is done by coupling 
the momentum and winding quantum numbers associated to the Euclidean time circle 
to the space-time fermion number $F$ \cite{AW, RK, AKADK}\cite{ckpt}.
The contributions of the right-moving world-sheet degrees of freedom to $F$ are
always even. Thus, at the level of the one-loop string partition function, the operation amounts to replacing the $\Gamma_{(1,1)}(R_0)$ lattice with 
\be 
\label{shiftLatT}
 \Gamma_{(1,1)}(R_0)\to \sum_{h_0,\tg_0}\Gamma_{(1,1)}[^{h_0}_{\tg_0}](R_0)~ 
(-)^{a\tg_0 + b h_0 + \tg_0 h_0}\, , 
\ee
where $\Gamma_{(1,1)}[^{h}_{\tg}]$ is a $\Z_2$-shifted lattice \cite{ShifftedLat} 
given by \cite{AW, RK, AKADK}\cite{ckpt}
\be
 \label{GammaLag} \Gamma_{1,1}[^h_{\tg}](R)  = {R\over \sqrt{\tau_2}}
\sum_{\tm,n} e^{-{\pi R^2\over \tau_2}|(2\tm + \tg) + (2n+h)\tau|^2}\, .
\ee

\noindent
The spontaneous breaking of space-time supersymmetry is done by coupling
the two $\Gamma_{(1,1)}$ lattices associated with the internal
circles of radii $R_4$ and $R_5$ to generic R-symmetry charges\cite{SSstringy,AW, RK, AKADK,ckpt}. 
In the case of models with $N_4=4$ initial supersymmetry, all such charges associated to
the left-moving world-sheet degrees of freedom are equivalent by symmetry. Different choices exist involving right-moving gauge R-charges \cite{ckpt}.
For example, consider the $E_8\times E_8$ models and decompose the
$E_8$ representations in terms of $SO(16)$ ones. 
One can choose R-charges which are odd for the $SO(16)$
spinorial representations and even for the others\footnote 
{In the $N_2=2$ orbifold models, one can choose
R-charges associated to the twisted $T^4$ planes to which the left-moving world-sheet degrees of
freedom contribute as well \cite{ckpt}.}.  

\noindent
We will present a class of cases, where starting with $N_4=4$, $E_8\times E_8$ vacua,
space-time supersymmetry is broken if we 
couple the $X^i$ lattices, $i=4,5$, to $F+\bar Q_i$, where the right-moving charges
$\bar Q_i$ are odd for the $SO(16)$ spinorial representations associated with one or both the 
$E_8$ factors.
So we replace the $X^4$ and $X^5$ lattices as follows:
\be
\label{shiftLatM}
\Gamma_{(1,1)}(R_i)\to \sum_{h_i,\tg_i}\Gamma_{(1,1)}[^{h_i}_{\tg_i}](R_i) ~ (-)^{(a+\bar Q_i)\tg_i + (b+\bar L_i)h_i +\epsilon_i\tg_i h_i}\, .
\ee
$\bar Q_i$, $i=4,5$, can be identified to be
either $\gamma$, $\gamma'$ or $\gamma+\gamma'$. [$\gamma$ and $\gamma^\prime$ are odd
for the corresponding $SO(16)$ spinorial representations.] $\bar L_i$ is equal to $\delta$, $\delta'$ or $\delta+\delta'$ respectively, as dictated by modular
invariance, and $\epsilon_i=0,1$ depending on the modular transformation $\tau\to
\tau+1$. Under this, 
\be a+\bar Q_i\to a+\bar Q_i\; , \qquad b+\bar L_i\to a+b+\bar Q_i+\bar L_i+\epsilon_i\, . \ee 
For instance, for $(\bar Q_i,\bar L_i)\equiv (\gamma+\gamma',\delta+\delta')$ one has
$\epsilon_i=1$, while for $(\bar Q_i,\bar L_i)\equiv (\gamma,\delta)$ or $(\gamma^\prime,\delta^\prime)$,  $\epsilon_i$ vanishes.
With these modifications taken into account, the one-loop string partition function is 
given by
\be
\label{Zeven}
\begin{array}{ll}
Z = &\!\! \displaystyle \frac{V_3}{(2\pi)^3}\int_\F {d^2\tau\over
2\tau_2^{5/2}}\ {\Gamma_{(4,4)}\over \eta^{12}\bar\eta^{24}} ~\frac 12 \sum_{a,b} (-)^{a+b+ab} \Th ab^4\\ 
  &\displaystyle \times \frac 14 \sum_{\gamma,\delta}\sum_{\gamma',\delta'}\bar\theta[^\gamma_\delta]^8 ~
\bar\theta[^{\gamma'}_{\delta'}]^8\prod_{i=0,4,5}~
\sum_{h_i,\tg_i}\Gamma_{(1,1)}[^{h_i}_{\tg_i}]~ (-)^{(a+\bar Q_i)\tg_i +(b+\bar L_i)h_i +\epsilon\tg_i
h_i} ~.
\end{array}
\ee
In this equation, we have $\bar Q_0=\bar L_0=0$.
Redefining $a=\hat a+\sum_i h_i$ and $b=\hat b+\sum_i \tg_i$, and using
the Jacobi identity \cite{ckpt}, one obtains
\be
 \label{Zodd}
\begin{array}{ll}
 Z=& \!\!\!\displaystyle - \frac{V_3}{(2\pi)^3}\int_\F \frac{d^2\tau}{2\tau_2^{5/2}}~{\Gamma_{(4,4)}\over \eta^{12}\bar\eta^{24}}
\sum_{h_i,\tg_i} \Th{1+\sum_i h_i}{1+\sum_i \tg_i}^4 ~ 
(-)^{\sum_i h_i + \sum_i \tg_i +(\sum_i h_i)(\sum_i \tg_i)}
\\
   &\displaystyle  \times~
\frac 14 \sum_{\gamma,\delta}\sum_{\gamma',\delta'}\bar\theta[^\gamma_\delta]^8 ~
\bar\theta[^{\gamma'}_{\delta'}]^8~
\prod_{i}\Gamma_{(1,1)}[^{h_i}_{\tg_i}]~
(-)^{\epsilon_ih_i\tg_i+\bar Q_i\tg_i+\bar L_i h_i}
\, .
\end{array}
\ee

\noindent
In the large radii regime $R_i \gg R_H$,
where $R_H$ is the Hagedorn radius,
 the system is free of tachyons. The odd winding sectors,
$h_i=1$, are exponentially suppressed. In this regime 
only the sectors $h_i=0$, $i=0,4,5$ and
$\tg_0 + \tg_4+\tg_5=1 \, {\rm modulo}\, 2$ contribute significantly
(the latter condition due to the fact that $\theta[^1_1]$ vanishes identically). 
Furthermore, if the internal lattice $\Gamma_{(4,4)}$ moduli are kept to be of order
unity, we can express the leading contributions as the following integral \cite{ckpt}:
\be
\label{Zdom}
\begin{array}{ll}
\nonumber Z=&\!\! \displaystyle \frac{V_3}{(2\pi)^3}\,R_0R_4R_5~ 
\sum_{\tg_i}{1-(-)^{\sum_i \tg_i}\over 2}~ \sum_s(-)^{\bar Q_4(s)\tg_4+ \bar Q_5(s)\tg_5} \\
&\displaystyle \times
\int_0^{\infty}
{d\tau_2\over \tau_2^4}\sum_{\tm_i}e^{-{\pi\over
\tau_2}[(2\tm_0+\tg_0)^2\,R_0^2 +
(2\tm_4+\tg_4)^2\,R_4^2+(2\tm_5+\tg_5)^2R_5^2]}\,.
\end{array}
\ee
In the first line the sum over $s$ runs over the $2^3 \times 504$ massless boson/fermion pairs of the  initially supersymmetric model. The contributions of massive states are exponentially suppressed,
of order $e^{-\pi
R_{i}}$.
The integral gives the pressure in the string frame: 
\be
\label{Znonthem}
\begin{array}{ll}
\displaystyle P_{\rm string}= {Z \over V_4} = &\!\!\displaystyle{R_4R_5\over (2\pi)^4}~{2\over \pi^3} \sum_{\tg_i}{1-(-)^{\sum_i\tg_i}\over 2}
~\sum_{s} (-)^{\tg_4 \bar Q_4(s)+ \tg_5 \bar Q_5(s)}\\
&\displaystyle \times ~\sum_{\tm_i}{1\over
\left[(2\tm_0+\tg_0)^2 R_0^2+(2\tm_4+\tg_4)^2 R_4^2 + (2\tm_5+\tg_5)^2R_5^2\phantom{\dot
\Phi}\!\!\!\!\!\right]^3} \, .
\end{array}
\ee

\noindent
We parameterize the various moduli as follows:
\be
\label{modT}
\begin{array}{cc}
\displaystyle T :={1\over 2\pi R_0\sqrt{\Re S}}\,\, ,\qquad  M:={1\over 2\pi \sqrt{ \T \Re S}}\,\,,\qquad 
 \Re T_1:={R_4R_5}\equiv \T\,\,,
 \\
\displaystyle \Re U_1:={R_5\over R_4}\equiv \U\,\,, \qquad u:= {R_0\over \sqrt{ \T}}\,\,,
\end{array}
\ee
where $S$ is the $4$d dilaton-axion modulus, $\Re S=e^{-2\phi_D}$. 
The two supersymmetry breaking scales, in the Einstein frame, are the temperature $T$ and the scale $M$, which 
for typical models determines the gravitino mass scale.
The pressure in this frame is related to the string frame pressure by:
\be
 \label{PT4}
P= \frac{1}{(\Re S)^2}\, P_{\rm string}=T^4 p(u,\U)\, ,
\ee
with
\be
p(u,\U)=n_{100}\,p_{100}(u,\U)+\,n_{010}p_{010}(u,\U)+n_{001}\,p_{001}(u,\U)+n_{111}\,p_{111}(u,\U)\
 .\ee
The coefficients $n_{\tg_0\tg_4\tg_5}$ are given in terms of the supersymmetry breaking R-charges, 
\be \label{nggg} n_{100} = 2^3\times 504, ~~~n_{010} = \sum_s (-)^{\bar Q_4(s)},~~~n_{001} = \sum_s
(-)^{\bar Q_5(s)},~~~n_{111} = \sum_s (-)^{\bar Q_4(s)+ \bar Q_5(s)}, \ee
while the dependence on the complex structure moduli $u$ and $\U$ involves the 
shifted Eisenstein functions:
\be \label{pggg} p_{\tg_0\tg_4\tg_5}(u,\U)={2\over \pi^3} \sum_{\tm_0,\tm_4,\tm_5}{u^4\over
\left[(2\tm_0+\tg_0)^2u^2+(2\tm_4+\tg_4)^2 \U^{-1} + (2\tm_5+\tg_5)^2\U\phantom{\dot
\Phi}\!\!\!\!\!\right]^3}\, .
\ee

%%%%%%%%%%%%%%%%%%%%%%%%%%%%%%%%%%%%%%%%

\subsection{Type II models}
\label{IIT}

We construct Type II models with similar thermal and supersymmetry breaking properties. In these examples, the internal manifold
involves either a $T^4$ factor for $N_4=8$ initial supersymmetry or a $T^4/\Z_2$ factor for $N_4=4$.
Orientifolds of these lead to models with $N_4=4$ and $N_4=2$ initial supersymmetry respectively, 
and include open string matter sectors.
At weak coupling in four dimensions, these are dual to heterotic models \cite{het-TIdual}, some of which we considered in the previous section.
Models with $N_4=2$ initial supersymmetry can also be constructed if we start with a $T^6/(\Z_2\times\Z_2^{\prime})$ orbifold
\cite{GKR}. 
We illustrate the derivation of the pressure in the intermediate cosmological region, with $T\ll T_H$, for the type II $N_4=4$ models, but the results can be generalized to the other cases.

\noindent
The $N_4=4$ partition function is 
\be
\begin{array}{ll}
Z = &\!\!\displaystyle {V_3\over (2\pi)^3}\int_\F {d^2\tau\over 2\tau_2^{5 / 2}}~{1\over (\eta\bar\eta)^8}
~\prod_{i=0,4,5}\Gamma_{(1,1)}(R_i)~{1\over 2}\sum_{H,G}Z_{(4,4)}[^H_G]
\\
&\!\! \displaystyle \times ~{1\over 2}\sum_{a,b}\theta[^a_b]^2\theta[^{a+H}_{b+G}]\theta[^{a-H}_{b-G}](-)^{a+b+ab}~{1\over 2}\sum_{\bar a,\bar b}\bar \theta[^{\bar a}_{\bar b}]^2\bar \theta[^{\bar a+H}_{\bar b+G}]\bar \theta[^{\bar a-H}_{\bar b-G}](-)^{\bar a+\bar b+\bar a\bar b}\,.
\end{array}
\ee
The $T^4/\Z_2$ part is given by \cite{ShifftedLat}:
\be
\begin{array}{lll}
\displaystyle Z_{(4,4)}[^H_G]={\Gamma_{(4,4)}\over (\eta\bar\eta)^4}\, ,&\when&(H,G)=(0,0)\\
\displaystyle Z_{(4,4)}[^H_G]={2^4\eta^2\bar\eta^2\over \theta [^{1-H}_{1-G}]^2\bar \theta [^{1-H}_{1-G}]^2}\, , &\when &(H,G)\neq(0,0)\,.
\end{array}
\ee
The characters  $H,G$ are integers defined modulo $2$.

\noindent
As usual, the finite temperature is implemented 
by inserting the thermal co-cycle and replacing the Euclidean time lattice 
as follows\cite{AW, RK, AKADK}:
\be
 \Gamma_{(1,1)}(R_0)\to \sum_{h_0,\tg_0}\Gamma_{(1,1)}[^{h_0}_{\tg_0}](R_0) 
~(-)^{(a+\bar a)\tg_0 + (b+\bar b) h_0}\, . 
\ee
In contrast to the heterotic case,
the contributions to the space-time fermion number $F$ from both the left-moving and
right-moving sectors can be odd or even. In the sequel, we denote by $F_L$ the
contribution of the world-sheet left-movers to the space-time fermion number and similarly
for $F_R$.

\noindent
There are several ways to break the initial $N_4=4$ supersymmetry spontaneously, either by
symmetric or asymmetric geometrical fluxes \cite{GeoFluxes,SSstringy,orbifolds}.
The two left-moving space-time supersymmetries can be broken 
if we couple either or both the $X^4$ and $X^5$ lattice charges to $F_L$, or
to left-moving R-charges associated with the twisted planes: $F_L+Q_i$. 
Also, the two right-moving space-time supersymmetries  
are broken by coupling the lattice charges to $F_R$ or to $F_R+\bQ_i$. 
Each lattice is replaced as follows \cite{akpt}: 
\be
\label{shiftLatB}
\Gamma_{(1,1)}(R_i)\to \sum_{h_i,\tg_i}\Gamma_{(1,1)}[^{h_i}_{\tg_i}](R_i) ~ (-)^{\left[(a+Q_i)\tg_i + (b+L_i)h_i +\tg_i h_i\right]\epsilon_i}~(-)^{\left[(\bar a+\bQ_i)\tg_i + (\bar b+\bL_i)h_i +\tg_i h_i\right]\bepsilon_i}\, ,
\ee
where $Q_i$, $\bQ_i$ can be set to zero or identified with the twist charge $H$.
Correspondingly
$L_i$, $\bar L_i$ can be set to zero or identified with the character $G$.
Also, we have introduced the parameters $\epsilon_i, \bepsilon_i$, taking the values 0 or 1, 
to indicate whether we couple the circle $i$ to the left- or right-movers.

\noindent 
In particular, we will examine 3 distinct cases where $N_4=4$ is spontaneously broken to $N_4=0$ (and then thermalized):

{\em -Case 1 :} Two asymmetric breakings, e.g. $(\epsilon_4, \bepsilon_4)=(1,0)$, $(\epsilon_5, \bepsilon_5)=(0,1)$. 

{\em -Case 2 :} One symmetric and one asymmetric breaking, e.g. 
$(\epsilon_4, \bepsilon_4)=(1,1)$, $(\epsilon_5, \bepsilon_5)=(0,1)$. 

{\em -Case 3 :} Two symmetric breakings, $(\epsilon_4, \bepsilon_4)=(1,1)$, $(\epsilon_5, \bepsilon_5)=(1,1)$. 

\noindent In addition, we will consider a case where the $N_4=4$ supersymmetry is partially  broken to $N_4=2$. The remaining supersymmetries are then broken by thermal effects:

 {\em -Case 1' :} Two left-moving asymmetric breakings, e.g. $(\epsilon_4, \bepsilon_4)=(1,0)$, $(\epsilon_5, \bepsilon_5)=(1,0)$. 

\noindent
The partition function can be written as follows:
\ba
\nonumber Z \!\!\!&=& \!\!\!{V_3\over (2\pi)^3}R_0R_4R_5\int_\F {d^2\tau\over 2\tau_2^4}
~{1\over (\eta\bar\eta)^8}~{1\over 2}\sum_{H,G}Z_{(4,4)}[^H_G] \\
 \label{ZevenB}
&&
 \!\!\! \times~ {1\over 2}\sum_{a,b}\theta[^a_b]^2\theta[^{a+H}_{b+G}]\theta[^{a-H}_{b-G}](-)^{a+b+ab}~{1\over 2}\sum_{\bar a,\bar b}\bar \theta[^{\bar a}_{\bar b}]^2\bar \theta[^{\bar a+H}_{\bar b+G}]\bar \theta[^{\bar a-H}_{\bar b-G}](-)^{\bar a+\bar b+\bar a\bar b} \\
\nonumber && \!\!\!\times \prod_{i=0,4,5}\left\{\sum_{h_i,\tg_i}\sum_{\tm_i,n_i}e^{-{\pi R_i^2\over \tau_2}|(2\tm_i+\tg_i) + (2n_i+h_i)\tau|^2}
(-)^{[(a+Q_i)\tg_i + (b+L_i)h_i +\tg_i h_i]\epsilon_i +[(\bar a+\bQ_i)\tg_i + (\bar b+\bL_i)h_i +\tg_i h_i]\bepsilon_i}\right\} .
\ea
Here $(\epsilon_0, \bar \epsilon_0)=(1,1)$ and $(Q_0, \bar Q_0)=(L_0, \bar L_0)=(0,0)$.

\noindent
As in the heterotic case, we are interested in the regime where the radii $R_i$, $i=0,4,5$, 
are much bigger than the Hagedorn radius, $R_i\gg R_H$. 
In this intermediate cosmological regime, the system is free of
any tachyonic instabilities. 
The odd winding sectors, $h_i=1$, are exponentially suppressed, and the only
significant contributions to the partition function occur 
for $h_i=0$, $\tg_0+\epsilon_4\tg_4+\epsilon_5\tg_5=1$ modulo $2$ and $\tg_0+\bepsilon_4\tg_4+\bepsilon_5\tg_5=1$
modulo $2$. 

\noindent
The pressure receives contributions from the untwisted sector, $H=0$, and from the 
twisted sector $H=1$.
In the untwisted sector, the result is given by 
\be
\label{ZBuntw}
\begin{array}{ll}
\displaystyle {Z_{\rm untwisted}\over V_4} = &\!\!\displaystyle{R_4R_5\over (2\pi)^4}~ \sum_{\tg_i}{1 - (-)^{\sum_i\epsilon_i\tg_i}\over 2}\; {1-(-)^{\sum_i\bar\epsilon_i\tg_i}\over 2}\\
\displaystyle &\displaystyle \times ~{2\over \pi^3}~\sum_{\tm_i}{n_0^{\mbox{\tiny untwisted}}\over \left[(2\tm_0+\tg_0)^2 R_0^2+(2\tm_4+\tg_4)^2 R_4^2 + (2\tm_5+\tg_5)^2R_5^2\phantom{\dot \Phi}\!\!\!\!\!\right]^3} \,.
\end{array}
\ee
$n_0^{\mbox{\tiny untwisted}}$ is the number of massless boson/fermion pairs in the untwisted sector
of the initially supersymmetric $N_4=4$ model. 
We have $n_0^{\mbox{\tiny untwisted}}=n_0/ 2$, where $n_0=2^7$ counts the massless pairs 
of the $N_4=8$ model; the factor of $1/2$ is due to the orbifolding. 

\noindent
In the twisted sector we have
\be
\label{ZBtw}
\begin{array}{ll}
\displaystyle 
{Z_{\rm twisted}\over V_4} =& \!\!\displaystyle {R_4R_5\over (2\pi)^4}~ \sum_{\tg_i}{1 - (-)^{\sum_i\epsilon_i\tg_i}\over 2}\; {1-(-)^{\sum_i\bar\epsilon_i\tg_i}\over 2}~(-)^{(\epsilon_4 Q_4+\bepsilon_4\bQ_4)\tg_4+(\epsilon_5 Q_5+\bepsilon_5\bQ_5)\tg_5}\\
&\displaystyle \times ~{2\over \pi^3}~\sum_{\tm_i}{n_0^{\mbox{\tiny twisted}}\over \left[(2\tm_0+\tg_0)^2 R_0^2+(2\tm_4+\tg_4)^2 R_4^2 + (2\tm_5+\tg_5)^2R_5^2\phantom{\dot \Phi}\!\!\!\!\!\right]^3} \, .
\end{array}
\ee
$n_0^{\mbox{\tiny twisted}}=2^8/2$ is the number of 
massless boson/fermion pairs in the twisted sector of the initially supersymmetric $N_4=4$ model. 
The $Q_4,\bQ_4,Q_5,\bQ_5$ appearing in Eq. (\ref{ZBtw}) can be either zero or identified to the twist charge $H=1$. 

\noindent
Using the definitions of the moduli introduced in Eq. (\ref{modT}), the
pressure $P$ is taking the same form as in Eq. (\ref{PT4}) with 
\be \label{pII}
p(u,\U)=\sum_{\mbox{\scriptsize $\begin{array}{c}
    \tg_0+\epsilon_4\tg_4+\epsilon_5\tg_5=1\,{\rm mod}\,2 \\ 
   \tg_0+\bepsilon_4\tg_4+\bepsilon_5\tg_5=1\,{\rm mod}\,2 \\ 
  \end{array}
$}}
n_{\tg_0\tg_4\tg_5} ~p_{\tg_0\tg_4\tg_5}(u,\U)\, ,
\ee
where the functions $p_{\tg_0\tg_4\tg_5}$ are given in Eq. (\ref{pggg}), and the coefficients
$n_{\tg_0\tg_4\tg_5}$ are similarly defined, in terms of supersymmetry breaking R-charges, as in Eq. (\ref{nggg}),
\be
\label{n8}
n_{\tg_0\tg_4\tg_5}= n_0= {2^8\over 2} ,~~~~~~~~~~~~~ ~~~~~~~~~~
~~~~~~~~~~~~~~ ~~~~~\mbox{for } N_4=8 \, ,
\ee
\be
\label{n4}
 n_{\tg_0\tg_4\tg_5}={n_0\over 2}\left(1+2(-)^{(\epsilon_4Q_4+\bepsilon_4\bQ_4)\tg_4+(\epsilon_5Q_5+\bepsilon_5\bQ_5)\tg_5}\right),~\mbox{for } N_4=4 \, .
\ee
In the $N_4=4$ cases, the
coefficients $n_{\tg_0\tg_4\tg_5}$ can take negative values as well. The results can be generalized to $N_4=2$ models.    

%%%%%%%%%%%%%%%%%%%%%%%%%%%%%%%%%%%%%%%

\subsection{The zero temperature limit}

Setting $T=0$, or $R_0 \to \infty$, in Eqs  (\ref{Znonthem}), (\ref{ZBuntw}) and (\ref{ZBtw}), we can obtain the one-loop effective potential at zero temperature. It arises from quantum effects due to the spontaneous breaking of supersymmetry.
For the heterotic models, the effective potential takes the form
\be
\label{VT=0}
\V=M^4\, v(\U),
\ee
where $M$ is defined in (\ref{modT}), and
\be
\label{vhet}
v(\U)=n_{10}v_{10}(\U)+n_{01}v_{01}(\U)\, .
\ee
The coefficients $n_{\tg_4\tg_5}$ are determined in terms of the R-charges,
\be
\label{nxx}
n_{\tg_4\tg_5}= \sum_s (-)^{\tg_4 \bar Q_4(s)+\tg_5 \bar Q_5(s)}\, ,
\ee
and
\be
v_{\tg_4\tg_5}(\U)=-{2\over \pi^3} \sum_{\tm_4,\tm_5}{1\over \left[(2\tm_4+\tg_4)^2 \U^{-1} + (2\tm_5+\tg_5)^2\U\phantom{\dot \Phi}\!\!\!\!\!\right]^3}\, .
\label{v}
\ee

\noindent
The type II effective potential takes a form similar to the heterotic one, as in Eq.  (\ref{VT=0}), where now
\be
\label{vII}
v(\U)=\sum_{\mbox{\scriptsize $\begin{array}{c}
    \epsilon_4\tg_4+\epsilon_5\tg_5=1\,{\rm mod}\,2 \\ 
   \bepsilon_4\tg_4+\bepsilon_5\tg_5=1\,{\rm mod}\,2 \\ 
  \end{array}
$}}
n_{\tg_4\tg_5} v_{\tg_4\tg_5}(\U)\,,
\ee
with $n_{\tg_4\tg_5}=n_{\tg_0\tg_4\tg_5}$, given in (\ref{n4}) (or (\ref{n8})).

%%%%%%%%%%%%%%%%%%%%%%%%%%%%%%%%%%%%

\subsection{Non thermal cosmologies}
\label{motion}

In the zero temperature limit, the 1-loop effective action takes the form:
\be
S=\int d^4x \sqrt{-{\rm det}~g}~\left\{{1\over 2} R-g^{\mu \nu}\left( {\partial_\mu S\partial_\nu \bar S\over (S+\bar S)^2} +    {\partial_\mu T_1\partial_\nu \bar T_1\over (T_1+\bar T_1)^2}+ {\partial_\mu U_1\partial_\nu \bar U_1\over (U_1+\bar U_1)^2}\right) -\V\right\}\, .
\ee
All other moduli can be frozen since they do not appear in the effective potential. More precisely, their contributions to the effective potential are exponentially suppressed. Freezing further $\Im S$, $\Im T_1$ and $\Im U_1$, 
we obtain for the Lagrangian
\be
\L={1\over 2} R-{1\over 2}\left( (\partial \phi_{ S})^2+ (\partial \phi_{\T})^2+ (\partial \phi_{\U})^2\right) -\,e^{-2\sqrt{2}(\phi_{ S}+\phi_\T)}\, {v(\U) \over (2\pi)^4}\, ,
\ee
where
\be
\label{phis}
\Re S := e^{\sqrt{2}\phi_{ S}}\, , \quad \T := e^{\sqrt{2}\phi_{\T}}\, , \quad \U := e^{\sqrt{2}\phi_{\U}}\, . 
\ee
It is useful to redefine the fields as follows
\be
\label{newphi}
\left(  \begin{array}{c}
   \phi \\ 
   \phi_- \\ 
  \end{array}
\right)
:=
\left(  \begin{array}{rr}
    -{1/\sqrt{2}}&-{1/\sqrt{2}} \\ 
   {1/ \sqrt{2}}& -{1/\sqrt{2}} \\ 
  \end{array}
\right)
\left(  \begin{array}{c}
  \phi_{ S}\\ 
   \phi_\T \\ 
  \end{array}
\right)
\ee
since the field $\phi_-$ does not appear in the potential: 
\be
\label{LT=0}
\L={1\over 2} R-{1\over 2}\left( (\partial \phi)^2+ (\partial \phi_-)^2+ (\partial \phi_{\U})^2\right)-M^4\, v(\U)\qquad \with \qquad M={e^{\phi}\over 2\pi}\, .
\ee

\noindent
We look for homogeneous and isotropic solutions where the metric 
is of the Friedmann-Robertson-Walker form, with vanishing spatial curvature: 
\be
\label{ds2}
ds^2=-N(t)^2dt^2+a(t)^2dx^i dx^i\, , \quad H\equiv \left({ \dot a \over a}\right)\, .
\ee
Here $N$ is the laps function, $a(t)$ the scale factor and $H$ the Hubble parameter.
In the gauge choice $N=1$, the gravitational field equations are
\ba
3H^2\!\!\!&=&\!\!\!{1\over 2}{\dot \phi}^2 + {1\over 2}{\dot \phi_-}^2+{1\over 2}{\dot \phi_\U}^2+\V\, \label{Hubble}\, , \\
2\dot H+3H^2\!\!\!&=&\!\!\!-{1\over 2}{\dot \phi}^2 - {1\over 2}{\dot \phi_-}^2-{1\over 2}{\dot \phi_\U}^2+\V\, . \label{eqaT0}
\ea
Their linear sum is independent of the fields kinetic terms:
\be
\dot H+3H^2=\V= M^4\,v(\U)\, .
\label{eqgrav}
\ee
Eq. (\ref{eqaT0}) follows 
by differentiating the Friedmann-Hubble Eq. (\ref{Hubble}), once the following moduli field equations are satisfied:  
\ba
\ddot \phi+3H\dot \phi\!\!\!&=&\!\!\!-{\partial \V\over \partial\phi}= - 4\,M^4\, v(\U)\, ,\label{eqphi}\\
\ddot \phi_\U+3H\dot \phi_\U\!\!\!&=&\!\!\!-{\partial \V\over\partial \phi_\U}= -\sqrt{2}\,\,M^4\, \U\,  v'(\U) \, , \label{eqU} \\
\ddot \phi_-+3H\dot \phi_-\!\!\!&=&\!\!\!0\, ,\label{eqphi-}
\ea
where prime derivatives are with respect to $\U$. The last equation can be integrated giving
\be
\label{c-}
{1\over 2}\, \dot \phi_-^2={c_-\over a^6}\, ,
\ee
where $c_-$ is a positive constant.
Eq. (\ref{eqU}) can be satisfied for a constant $\U$, if there exists a solution to
\be
\label{min}
v'(\U)=0\, .
\ee
We will look for models for which this extremum is a local minimum so that 
the complex structure modulus $\U$ is stabilized.
The compatibility of Eqs  (\ref{eqgrav}) and (\ref{eqphi}) requires that there exists a constant $c_\phi$ such that:
\be
\dot \phi=-4H+{c_\phi\over a^3}\, .
\ee
In \cite{BKP}, it is shown that the solution for  $c_\phi=0$ is an attractor. 
Thus, we concentrate on the case $c_\phi=0$, so that 
\be
M\equiv {e^\phi\over 2\pi}=M_0\left( {a_0\over a}\right)^4\, ,  
\ee
where $M_0a_0^4$ is a positive integration constant. 

\noindent
Higher loop contributions in the effective action are suppressed by powers of 
${\rm {Re}}~S \sim e^{\phi}$. In particular, they remain suppressed at future times as
long as the scale factor $a$ is growing. Thus the system remains weakly coupled in the future.
If we extrapolate the solution arbitrarily back in time, 
we encounter in many of these models tachyonic instabilities when
the supersymmetry breaking scale $M$ becomes of order the string scale, before the Big Bang. Eventually perturbation theory
breaks down. The analysis of the dynamics
of tachyon condensation in this highly stringy regime is beyond the scope of this work. We also point out that in certain asymmetric 
orbifold constructions, despite the breaking of supersymmetry, the spectrum is free of tachyons for any value of the radial moduli \cite{akpt, ACI}.
It would be interesting to analyze cosmological solutions in such type of models.     

\noindent
The Hubble equation takes the form:
\be
\label{Hubble16}
3H^2=-{c_m\over a^{6}}+{c\over a^{16}}\qquad \mbox{where}\qquad c_{m}={3\over 5}\, c_->0 \; ,\quad c=-v(\U)\times {3\over 5}\, M_0^4\, a_0^{16}.
\ee
If $c>0$, one has for $c_m=0$
\be
a(t)=A\;  t^{1/8}\quad \mbox{where}\quad A=2^{3/8}\left(c\over 3\right)^{1/16}\, .
\ee
When the kinetic energy for $\phi_-$ is switched on, i.e. when $c_m>0$, 
one has a big bang/big crunch cosmology.
The solution $t(a)$ is given by
\be
t(a)=\pm \, t_0 \int_{a/A}^1{x^7dx\over \sqrt{1-x^{10}}}\, , \quad 0\leq a\leq A \quad \mbox{where}\quad A=\left({c\over c_m}\right)^{1/10}\; , \quad t_0=\sqrt{3}c^{3/10}c^{-4/5}\, .
\ee

\noindent
We investigate whether some of the heterotic and type II models 
we considered satisfy the minimization condition (\ref{min}), which fixes the modulus $\U$,  
and the positivity of the parameter $c$, Eq. (\ref{Hubble16}), which
allows for real time solutions:
\be
\label{cond}
\mbox{\em Extremum : }v'(\U)=0\; , \qquad \mbox{\em Stability : }v^{\prime \prime}(\U)>0\; ,\qquad\mbox{\em Real time : } v(\U)<0\, .
\ee
The shape of the potential as a function of $\U$ depends on the R-symmetry breaking charges which define the coefficients $n_{\tg_4\tg_5}$. Since the functions $v_{\tg_4\tg_5}$ in Eq. (\ref{v}) are negative, 
we need some of the $n_{\tg_4\tg_5}$ to be positive. 

\begin{figure}[h!]
\begin{center}
\vspace{.3cm}
\includegraphics[height=4cm]{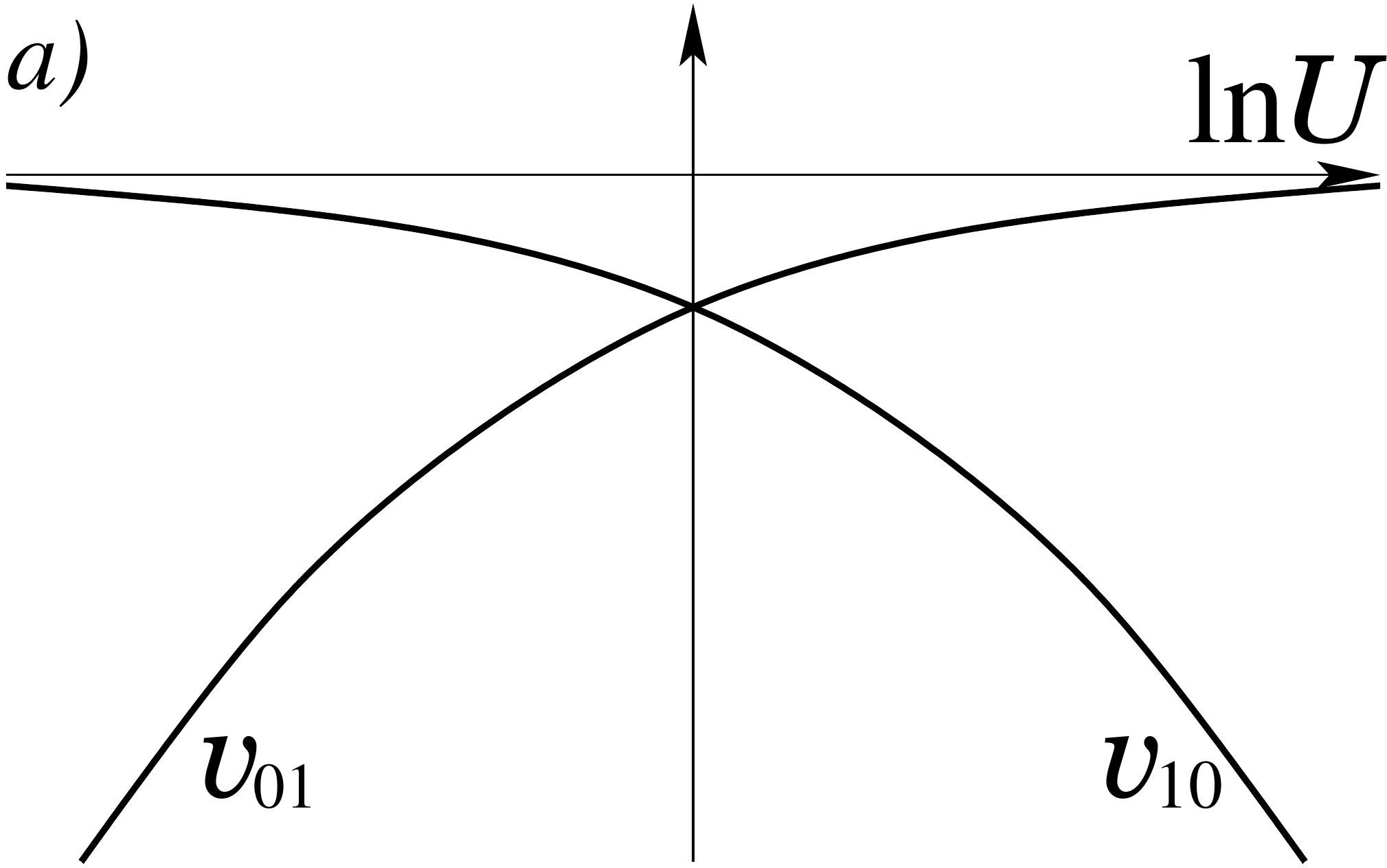}
\qquad\qquad 
\includegraphics[height=4cm]{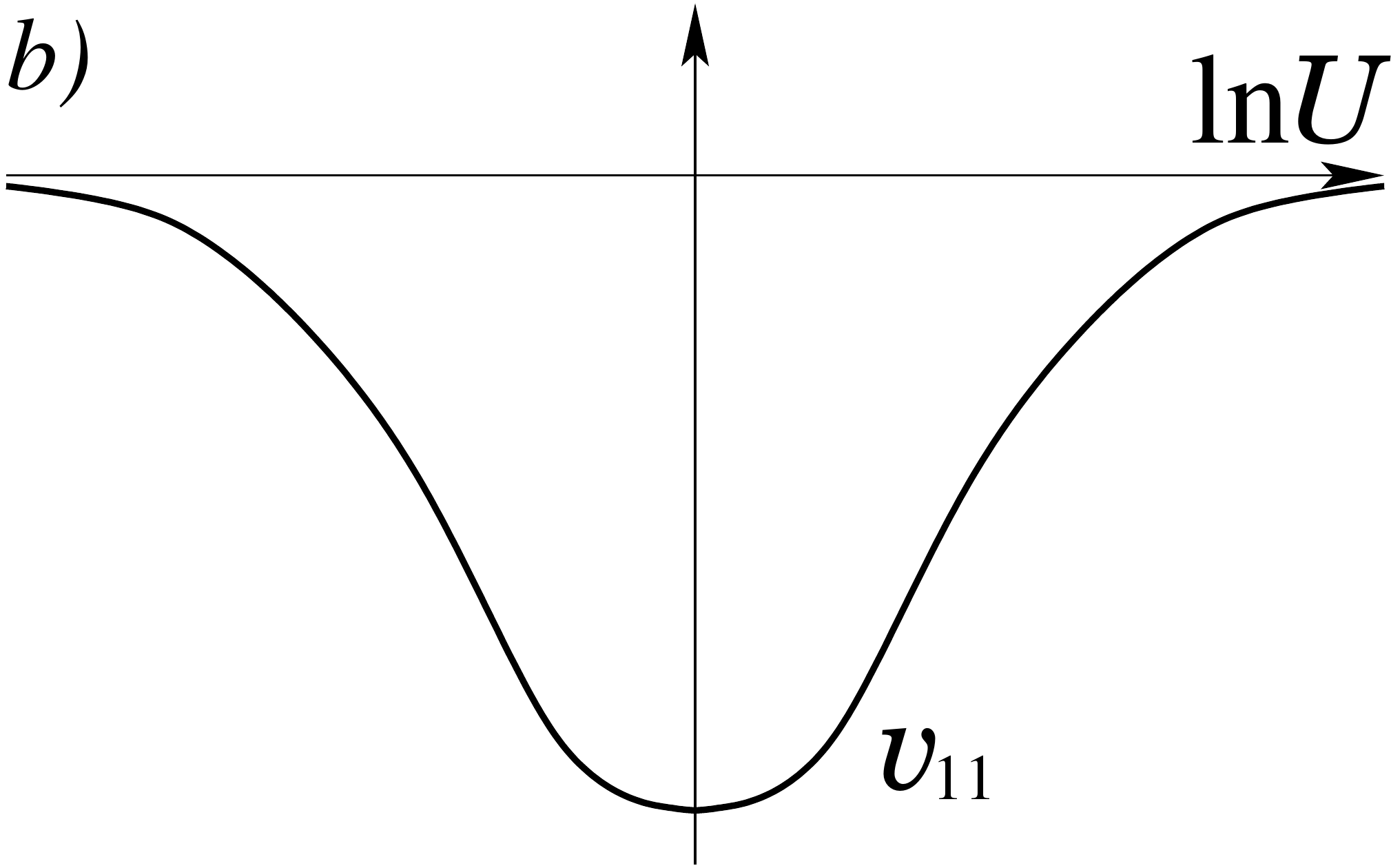}
\caption{\footnotesize \em The functions $v_{\tg_4\tg_5}$ versus $\ln \U$.}
\vspace{-.5cm}
\label{vg4g5}
\end{center}
\end{figure}

\noindent 
In the heterotic models, the functions $v_{10}$ and $v_{01}$ defining the effective potential, Eq. (\ref{vhet}), are monotonic functions of $\U$, the former decreasing and the latter increasing, (see Fig. \ref{vg4g5}{\em a}). 
To have an extremum, $n_{10}$ and $n_{01}$ must be of the same sign. 
The real time condition requires this sign to be positive, 
which implies that the solution is unstable under fluctuations of $\U$. Stable cosmological solutions exist only at non-zero temperature, as we will see in the next section. We may have, however, stationary domain wall solutions.

\noindent 
We now study the effective potential given in Eq. (\ref{vII}) 
for the different type II cases.

{\em -Case 1 :} The potential is proportional to the function $v_{11}$, 
which is 
invariant under $\U\to 1/\U$.  This implies that a stationary point, which is a minimum, occurs at $\U=1$, (see Fig. \ref{vg4g5}{\em b}). 
The conditions (\ref{cond}) are simultaneously satisfied if $n_{11}>0$. 
This can be realized by choosing $Q_4=\bQ_5= 0$ or $H$.

{\em -Case 2 :} Since the potential is proportional to the monotonic function $v_{01}(\U)$, 
there is always a runaway behavior for this model: 
$\U\to 0$ \ie $R_4\to +\infty$ for $n_{01}>0$ and  $\U\to +\infty$ \ie $R_5\to +\infty$ for $n_{01}<0$. Thus, the dynamics drives the system out of the scope of our analysis, and one should look for a  solution in five dimensions, where supersymmetry is spontaneously broken by the remaining finite size internal radius. 

{\em -Case 3 :} This type II model has an effective potential of the form encountered in the heterotic case.

{\em -Case 1' :} The model is supersymmetric with a flat effective potential.

\noindent At finite temperature, we are going to see that the situations in the type II cases also change drastically.

%%%%%%%%%%%%%%%%%%%%%%%%%%%%%%%%%%%%%%%%
%%%%%%%%%%%%%%%%%%%%%%%%%%%%%%%%%%%%%%%%

\section{Thermal cosmologies}
\label{finiteT}

\noindent As we already stated before, we work in a weak coupling regime, with the temperature and supersymmetry breaking scales 
sufficiently below the Hagedorn temperature.  It is very important to stress here, that the string perturbative computations we use to obtain the energy density and pressure are free of any ultraviolet and infrared ambiguities and the contributions of moduli that are not participating in the supersymmetry breaking mechanism are exponentially suppressed and can be consistently neglected. The solutions we obtain describe the cosmological evolution at times much after a Hagedorn phase 
transition, from a high temperature stringy phase. The perturbative analysis is not sufficient to analyze the dynamics of the system at the transition, 
but we expect the solutions to match the correct behavior of the system when the Universe has cooled down at temperatures below 
the Hagedorn temperature. In \cite{BKP}, it is explicitly shown that for generic initial conditions after the Hagedorn exit, the solution is always attracted to the
ones we display below.

%%%%%%%%%%%%%%%%%%%%%%%%%%%%%%%%%%%%%%%%

\subsection{Equations of motion and thermodynamics}

\noindent Using the field redefinitions (\ref{phis})--(\ref{newphi}) and the FRW ansatz (\ref{ds2}), the thermal effective action is
\be
\label{ST}
S=-{1\over 6}\int dt \, Na^3\left({3\over N^2}\, H^2 -{1\over 2N^2}\, \dot \phi^2-{1\over 2N^2}\, \dot \phi_-^2-{1\over 2N^2}\, \dot \phi_\U^2-P\right)  \, ,
\ee
where the pressure can be expressed in terms of the supersymmetry breaking scale $M(\phi)$ as
\be
\label{defM}
P= M^4\, {p(u,\U)\over u^4}\, , \qquad M= {e^{\phi}\over 2\pi}\, , \qquad u= {M\over T}\, .
\ee
The gravity equations are obtained by varying with respect to $N$ and $a$, 
\be
\label{geneFried}
{3\over N^2}\, H^2={1\over 2N^2}\, {\dot \phi}^2+{1\over 2N^2}\, {\dot \phi_-}^2+{1\over 2N^2}\, {\dot \phi_\U}^2+\rho \qquad \with \qquad \rho:=-P-N\, {\partial P\over \partial N}\, ,
\ee
\be
\label{genevara}
{2\over N^2}\, \dot H+{3\over N^2}\, H^2-{2\over N^3}\,H\dot N=-{1\over 2N^2}\, {\dot \phi}^2-{1\over 2N^2}\, {\dot \phi_-}^2-{1\over 2N^2}\, {\dot \phi_\U}^2 - P-{1\over 3}\, a\, {\partial P\over \partial a}\, ,
\ee
where in general $P$ can depend on $N$ and $a$. To determine the dependence, we recall in which specific frame we computed $P$:
\be
P={Z\over (\Re S)^2V_4}={Z\over N\,  a^3}\qquad \Longrightarrow \qquad N=2\pi R_0\sqrt{\Re S}= {1\over T}\; , ~~a=2\pi R\sqrt{\Re S}\, ,
\ee
where $R_1=R_2=R_3\equiv R$ are the radii of the large three spatial directions (before the large volume limit $R\to +\infty$ is taken). 
Since $P=T^4 p(u,\U)$ and $T$ is identified with the inverse of the laps function, we have
\be
\label{rhoP}
\rho=T~{\partial P\over \partial T}-P\qquad \and\qquad {\partial P\over \partial a}=0\, .
\ee
It is remarkable that these expressions are identical to the ones derived from thermodynamics. 
We thus show that the variational principle is in perfect agreement with thermodynamics. 

\noindent 
Since  $P$ (in the action (\ref{ST})) and $\rho$ (in Eq. (\ref{geneFried})) are scalars under time reparameterizations, 
we can write the gravitational equations in the simple gauge $N= 1$ as follows:  
\ba
\label{HubbleT}
3H^2\!\!\!&=&\!\!\!{1\over 2}{\dot \phi}^2+{1\over 2}{\dot \phi_-}^2+{1\over 2}{\dot \phi_\U}^2+\rho \, ,\\
2\dot H+3H^2\!\!\!&=&\!\!\!-{1\over 2}{\dot \phi}^2-{1\over 2}{\dot \phi_-}^2-{1\over 2}{\dot \phi_\U}^2 - P\, .
\label{eqaT}
\ea
Combining Eqs  (\ref{PT4}) and (\ref{rhoP}), we obtain
\be
\rho=T^4r(u,\U)\qquad \mbox{where}\qquad r=3p-u\partial_u p\, .
\label{rp}
\ee
The equations of motion for the moduli fields can be written as follows: 
\ba
\ddot \phi+3H\dot \phi\!\!\!&=&\!\!\!{\partial P\over \partial\phi}\equiv T^4\left\{ 3p(u,\U)-r(u,\U)\right\} \, ,\label{eqphiT}\\
\ddot \phi_\U+3H\dot \phi_\U\!\!\!&=&\!\!\!{\partial P\over\partial \phi_\U}\equiv T^4 \,\sqrt{2}\,\,\U\,\partial_\U p(u,\U) \, ,\label{eqphiU} \\
\ddot \phi_-+3H\dot \phi_-\!\!\!&=&\!\!\!0\, ,
\ea
 where the r.h.s. of Eq. (\ref{eqphiT}) follows from Eqs  (\ref{defM}) and (\ref{rp}). 

\noindent
The last equation gives (\ref{c-}). We would like to find solutions to the remaining system of equations, with  $\U$ stabilized. Then, Eq. (\ref{eqphiU}) amounts to the following algebraic equation 
 \be
 \label{eq1}
D(u,\U):= \,\U\,\partial_\U p =0\, ,
 \ee
requiring that $u$ is also a constant. It follows that the time dependence of $\rho$ and $P$ arises from the $T^4$ pre-factors only. From the relations in (\ref{defM}), we have that  
\be
\label{phiT}
M(\phi)=u T\qquad \Longrightarrow \qquad \dot \phi={\dot T\over T}\, .
\ee
Instead of solving the scale factor Eq. (\ref{eqaT}), we choose to solve the equation that arises from the conservation of the energy-momentum tensor:
\be
\label{CE}
{d \over dt}\left({1\over 2} \dot \phi^2+{1\over 2} \dot
  \phi_\U^2+{1\over 2} \dot
  \phi_-^2+\rho\right)+
3H\left(\dot\phi^2+ \dot
  \phi_\U^2+\dot\phi_-^2+\rho+P\right)=0\, .
\ee
Using the equations of motion for the scalar fields and (\ref{phiT}), this gives
\be
\label{phiH}
\dot \phi={\dot T\over T}=-H\qquad \Longrightarrow \qquad a\, T=a_0\,T_0\, ,
\ee
where $a_0T_0$ is a positive integration constant. 

\noindent 
Next, we consider the linear sum of Eqs  (\ref{eqaT}) and (\ref{HubbleT}),
\be
\dot H+3H^2={1\over 2}(\rho-P)={1\over 2}T^4\left\{r(u,\U)-p(u,\U)\right\}\,. 
\label{eqgravT}
\ee
Using (\ref{phiH}), the compatibility between this equation and (\ref{eqphiT}) 
 implies the following thermal equation of state: 
\be
\label{special}
\rho=5P\, .
\ee
This is one of the main results of this paper. It reminds us of the analogous equation, $\rho=4P$, derived in \cite{ckpt,ak}, when a single modulus was participating in the spontaneous supersymmetry breaking mechanism. These results are very suggestive, and we conjecture that they will be generalized to the cases when more moduli participate in the supersymmetry breaking. When $n$ such fields are involved, we expect the equation of state to take the form:
\be
\rho=(3+n)P\,.
\ee 

\noindent
In the two moduli case, which we are considering here, Eq. (\ref{special}) can also be written as 
\be
\label{eq2}
C(u,\U):=(2+u\partial_u)p = 0\, . 
\ee
As a result, the complex structure ratios $(u,\U)$ are determined by the equations $D=C=0$.
It is interesting that along this critical trajectory, complex structure moduli participating in the breaking of supersymmetry are stabilized, and thus the cosmology is
characterized by a single running scale.

\noindent The time dependence of the scale factor is dictated by the Friedmann-Hubble Eq. (\ref{HubbleT}). The latter takes the form 
\be
\label{Hubble4}
3H^2={c_r\over a^4}+{c_m\over a^6}\qquad \mbox{where}\qquad c_r=6(a_0T_0)^4p(u,\U)\; , \quad c_m={6\over 5}\, c_->0\, .
\ee
When $c_m=0$, the universe is effectively radiation dominated and a cosmological solution exists if the constant $p(u,\U)$ is positive:
\be
a(t)=B\sqrt{t}\qquad \mbox{with}\qquad B=\sqrt{2}\left({c_r\over 3}\right)^{1/4}\, .
\ee

\noindent
When $c_m$ is non-trivial,  the time $t$ can be expressed as a function of the scale factor as follows:
\be
\begin{array}{lll}
 \mbox{if $c_r>0$ : }& \displaystyle t(a)=t_0 \int^{a/B}_0{x^2dx\over \sqrt{1+x^2}}\; , &\forall a\ge 0 \, ,\\
\mbox{if $c_r<0$ : }& \displaystyle t(a)=\pm \, t_0 \int_{a/B}^1{x^2dx\over \sqrt{1-x^2}}\, , & 0\leq a\leq B \, ,
\end{array}
\ee
where 
\be
B=\sqrt{{c_m\over |c_r|}}\qquad \mbox{and}\qquad t_0=\sqrt{3}c_m |c_r|^{-3/2}\, .
\ee
In the explicit examples presented in the following section, we always find $c_r>0$. It would be interesting to find if  models with ``negative effective radiation energy density'', $c_r/a^4$, are allowed. 

\noindent
As noticed in \cite{ckpt}, the fact that the cosmological evolution we have found behaves effectively like a four-dimensional universe filled with thermal radiation is not in contradiction with the state equation $\rho=5P$. The reason is that the total energy density and pressure contain the ``cold" part associated to the kinetic energy of $\phi$. When $c_m=0$, one has $\dot \phi^2/2=\rho/5$, so that
\ba
\nonumber \rho_{\rm tot}&=&{1\over 2}\dot \phi^2+\rho={6\over 5}\, \rho\\
P_{\rm tot}&=&{1\over 2}\dot \phi^2+P={2\over 5}\, \rho\, ,
\ea
in agreement with the expected state equation $\rho_{\rm tot}=3P_{\rm tot}$.

%%%%%%%%%%%%%%%%%%%%%%%%%%%%%%%%%%%

\subsection{Some stringy examples}
\label{solT}

We examine whether the extremization condition (\ref{eq1}) and the compatibility condition (\ref{eq2}) are simultaneously satisfied in the various heterotic and type II models under consideration:
\be
\label{condT}
\mbox{\it Extremum : }D(u,\U)=0\; , ~\mbox{\it Compatibility : } C(u,\U)=0\; ,~\mbox{\it Stability : }\partial_\U^2p(u,\U)<0\, .
\ee
As in the non-thermal situation,
the shape of the potential depends on the R-symmetry breaking charges. Their choices determine the coefficients $n_{\tg_0\tg_4\tg_5}$ that satisfy $-n_{100}\leq n_{\tg_0\tg_4\tg_5}\leq n_{100}$.
 
\noindent
However, in the heterotic cases, $n_{111}$, $n_{010}$ and $n_{001}$ are not totally arbitrary.  
It is convenient to parameterize the a priori allowed models by separating the $n_{100}$ states into 4 groups, depending on their parity under the operators $(-)^{\bar Q_4}$ and $(-)^{\bar Q_5}$, as shown in Table \ref{abc}. 
\be
\label{abc}
  \begin{array}{|c||c|c|c|c|}
  \hline
    & n_{100} \xi_1 \mbox { states}&   n_{100}\xi_2  \mbox { states}&   n_{100}\xi_3 \mbox { states} &  n_{100} (1-\xi_1-\xi_2-\xi_3) \mbox { states} \\ \hline\hline
(-)^{\bar Q_4} & + & + & - & - \\ \hline
(-)^{\bar Q_5} & + & - & + & - \\ \hline
  \end{array}
\ee
We observe that the parameter space of models is the tetrahedron: 
\be
\left\{ (\xi_1,\xi_2,\xi_3)\in [0,1]^3\quad \mbox{such that} \quad \xi_1+\xi_2+\xi_3\le 1\phantom{\dot \Phi}\!\!\!\!\right\}\, ,
\ee
which constrains the ratios:
\be
r_{010}:={n_{010}\over n_{100}}=2(\xi_1+\xi_2)-1\, , ~ r_{001}:={n_{001}\over n_{100}}=2(\xi_1+\xi_3)-1\, , ~ r_{111}:={n_{111}\over n_{100}}=1-2(\xi_2+\xi_3)\, .
\ee
The conditions (\ref{condT}) and the $\xi_2 \leftrightarrow \xi_3$ duality symmetry can be visualized geometrically in terms of the tetrahedron representation. Some type II models are also characterized in terms of this representation.

\begin{figure}[h!]
\begin{center}
\vspace{.3cm}
\includegraphics[height=5cm]{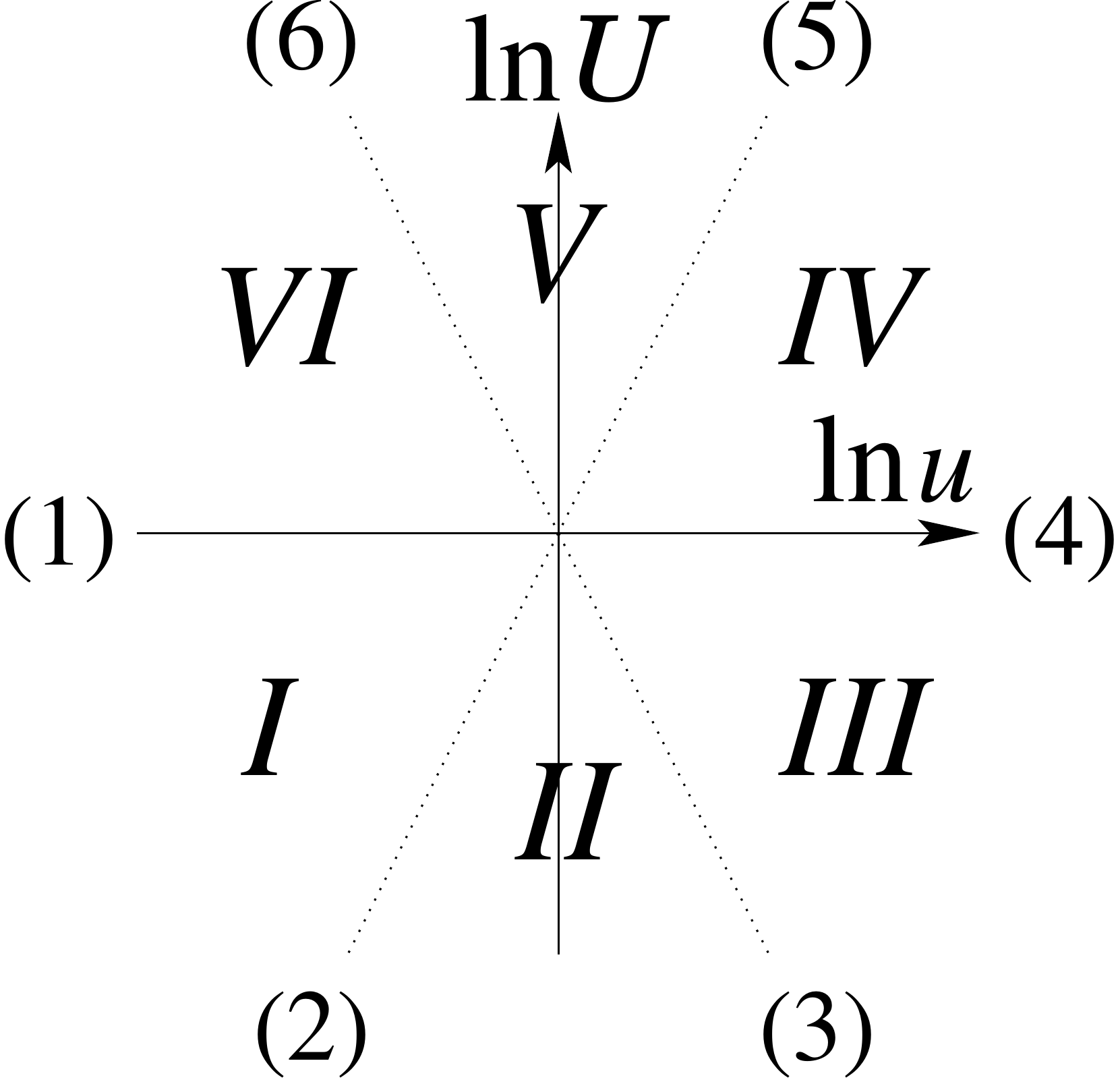}
\caption{\footnotesize \em The $(\ln u,\ln \U)$-complex plane can be divided in $6$ sectors, $I,\dots,VI$, separated by 6 edges of slope $-2$, $0$ or $2$. The power expansions of $p(u,\U)$, which are listed in the Appendix are well defined in each sector.}
\vspace{-.5cm}
\label{sectors}
\end{center}
\end{figure}

\noindent
 In the large/small $u$ and $\U$ regimes, $p$ contains exponentially suppressed contributions that we have to neglect by consistency. The dominant contributions take the form of a linear sum of a finite number of monomials $u^a\U^b$.  
The $(\ln u, \ln\U)$-plane is divided into 6 sectors inside of which a power expansion is defined, (see Fig. \ref{sectors}). The boundaries of these sectors are the lines $\U=\lambda u^\omega$, where $\omega=-2,0,2$. 
We show in the Appendix the power expansion of $p(u,\U)$ in each sector. More accurate expressions  are also given along the lines $\U=\lambda u^{\pm 2,0}$ that are useful to connect the power expansions on each sides. $D(u,\U)=0$ and $C(u,\U)=0$ define curves which asymptote to the lines $\U=\lambda u^{\pm 2,0}$, for which we have determined the intercept $\lambda$ as a function of the ratios $r_{\tg_0\tg_4\tg_5}$. 
The constraints are simultaneously satisfied if these curves meet at a point $(u_c,\U_c)$. Then,
the stability condition, $\partial_\U^2 p(u_c,\U_c)<0$, and the sign of the radiation density are determined.
The $C(u,\U)=0$ constraint requires to have at least one negative $r_{\tg_0\tg_4\tg_5}$. 

\begin{figure}[h!]
\begin{center}
\vspace{.3cm}
\includegraphics[height=4cm]{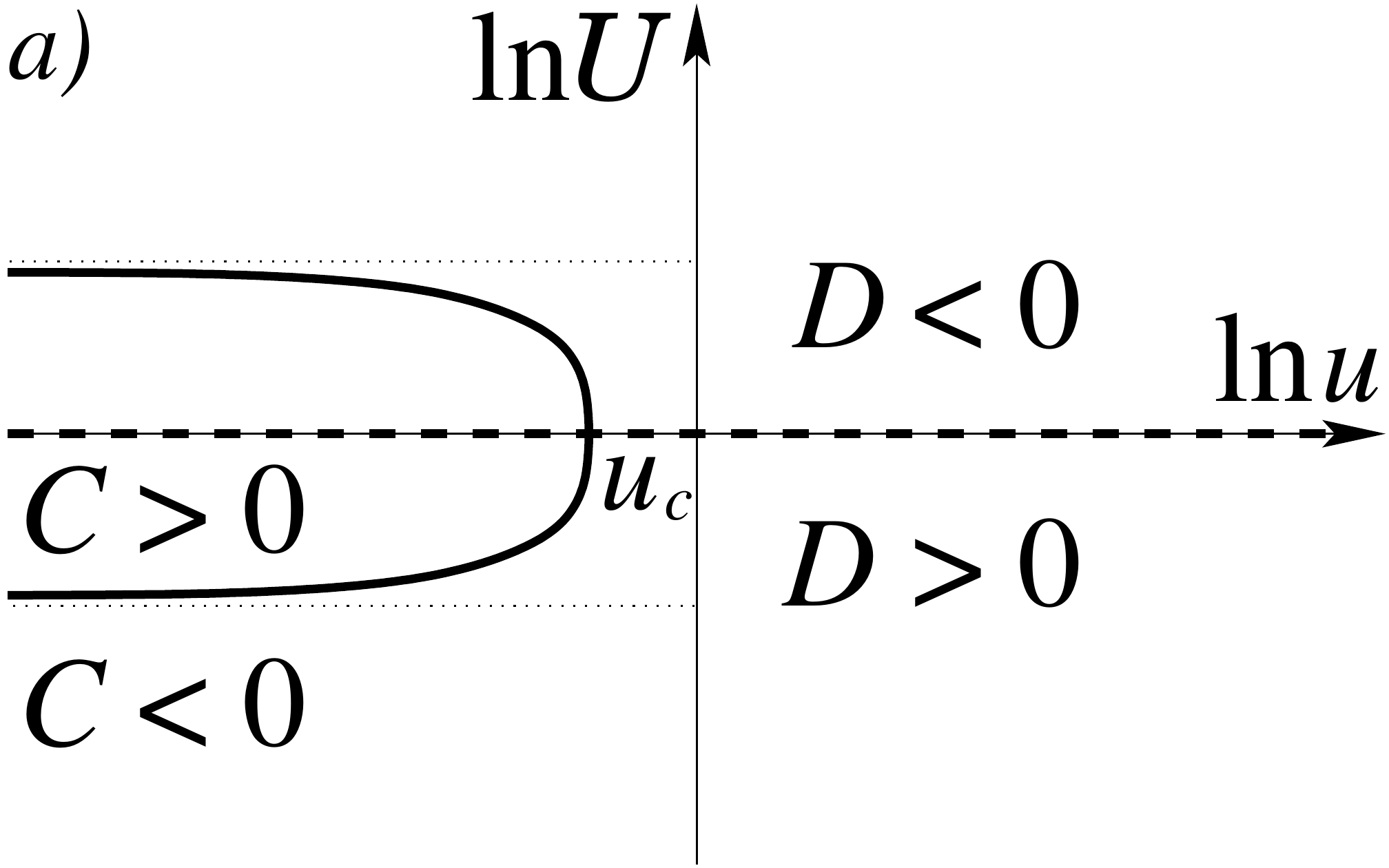}
\qquad\quad
\includegraphics[height=4cm]{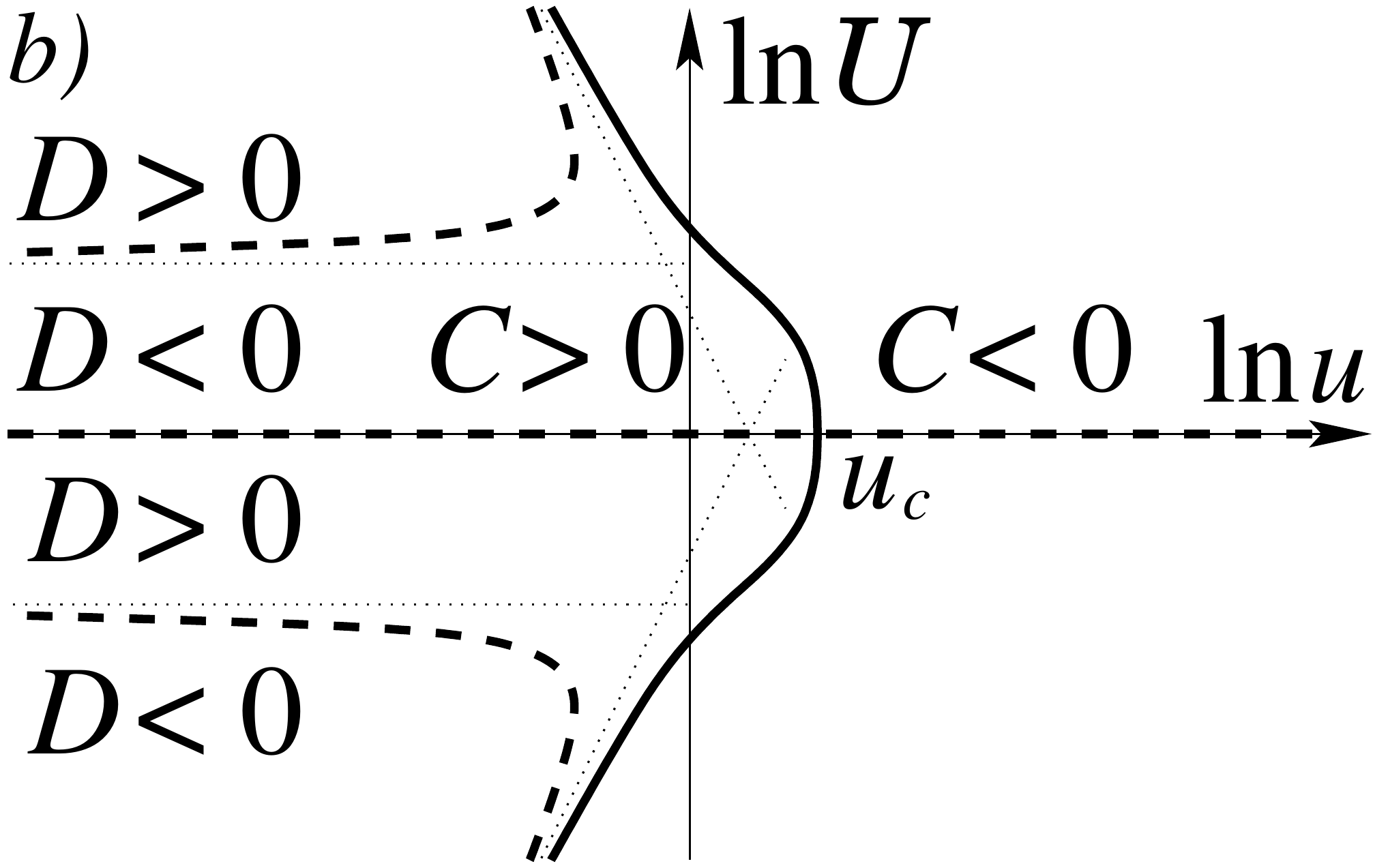}
\caption{\footnotesize \em For a pressure of the form (\ref{case3A}), the curves defined by $C(u,\U)=0$ (straight lines) and $D(u,\U)=0$ (dashed lines) are represented. When $r_{c3}<r<r_{c2}$ (Fig. a), there is a stable cosmological solution $(u_c,\U_c=1)$.  It corresponds to a global minimum of the thermal effective potential. When $r_{c2}\le r<0$ (Fig.  b), there is a stable cosmological solution $(u_c,\U_c=1)$ that corresponds to a local minimum. Two run away behaviors that bring the system to five dimensions are also allowed.}
\vspace{-.5cm}
\label{case_3A}
\end{center}
\end{figure}

%%%%%%%%%%%%%
\noindent  
$\bullet$ Our analysis shows the existence of non-trivial thermal cosmological solutions in heterotic models with $\bar Q_4\equiv \bar Q_5$, and type II {\em Case 3} models with $Q_4+\bar Q_4\equiv  Q_5+\bar Q_5$, where 
 the pressure takes the general form
\be
\label{case3A}
p(u,\U)= n_{100}[p_{100}+r (p_{010}+p_{001})+p_{111}] \, ,
\ee
with $-1\leq r\leq 1$. These models lie along the edge $\xi_2=\xi_3=0$ of the tetrahedron. The duality symmetry $\U\to \U^{-1}$ implies that along the axis $\U\equiv 1$,  $D(u,\U)=0$.  As $r$ varies, we find 4 distinct patterns, specified by
\be
r_{c3}=-{f^{ee}_{5/2}(1)+f^{oo}_{5/2}(1)\over  2 f^{oe}_{5/2}(1)}\simeq -0.215\; , ~~~r_{c2}=-{S^e_5\over S^o_5}=-{1\over 31} \;  ~{\rm and}~~r_{c1}=0\, ,
\ee
where the functions in the definition of $r_{c3}$ can be found in  Eq. (\ref{ffunc}). \\
- It turns out that there are no cosmological solutions with constant $u$ and $\U$ when $r<r_{c3}$ or $r>0$, since then $C(u,\U)\ne 0$ everywhere. \\
- A stable cosmological solution exists when $r_{c3}< r<r_{c2}$, with $c_r>0$, (see Fig. \ref{case_3A}{\em a}). It corresponds to  a global minimum of the thermal effective potential.
\\
\noindent - When $r_{c2}\le r<0$, a stable cosmological solution with $\U_c=1$ still exists, with $c_r>0$, but this corresponds to a local minimum of the thermal effective potential. Actually, new branches of the locus $D=0$ are present and converge exponentially towards the curve $C=0$, (see Fig. \ref{case_3A}{\em b}). Formally, their common asymptotes define flat directions. However,  since these ``solutions'' imply $R_5$ (or $R_4$) to be very large, they are out of the scope of our analysis. They are better understood in terms of runaway behaviors that decompactify the system to five dimensions, where supersymmetry is broken by the remaining finite size internal radius $R_4$ (or $R_5$) and thermal effects.\footnote{
A similar runaway behavior can be realized in type II {\em Case 2} models, with generic operator $\bar Q_5$. They involve a pressure $p(u,\U)= n_{100}[p_{100}+r p_{001}]$, where $-1\leq r\leq 1$. Their representative points in the tetrahedron satisfy $\xi_1=\xi_3={1\over 2}-\xi_2$, where $0\le \xi_2\le {1\over 2}$. When $r$ varies, there is a phase where the curves $D=0$ and $C=0$ are non-trivial and asymptotic to one another. The situation is similar to what is observed in the lower half plane of Fig.  \ref{case_3A}{\em b}.
}

\noindent
For the heterotic model with the choice $\bar Q_4=\bar Q_5=\gamma+\gamma'$, one has
\be
\label{nhet}
\begin{array}{lll}
n_{100}=n_{111}\!\!\!&=&\!\!\!2^3\times 504\,,\\
 n_{010}=n_{001}\!\!\!&=&\!\!\!2^3\left[~[2]_{X_{2,3}}+[6]_{T^6}+[120-128]_{E_8}+[120-128]_{E_8^{\prime}}~\right] =-2^3\times 8 \, ,
 \end{array}
 \ee
so that $r_{c2}<r=-1/63<r_{c1}$. In this specific case, the cosmological solution corresponds to $(u_c,\U_c)\simeq (1.649, 1)$, where $c_r\simeq  0.0708 \times {6(a_0T_0)^4}$.
 
\noindent Another model considered in \cite{ckpt} is based on the heterotic $T^4/\Z_2$
orbifold, with non-Abelian gauge group $E_8\times E_7\times SU(2)$. 
In that case, the parameter $r$ satisfies $r_{c2}<r=-1/127<r_{c1}$ and the corresponding cosmological solution fixes $(u_c,\U_c)\simeq (1.996, 1)$, with $c_r\simeq  0.0762  \times {6(a_0T_0)^4}$. 

\begin{figure}[h!]
\begin{center}
\vspace{.3cm}
\includegraphics[height=4cm]{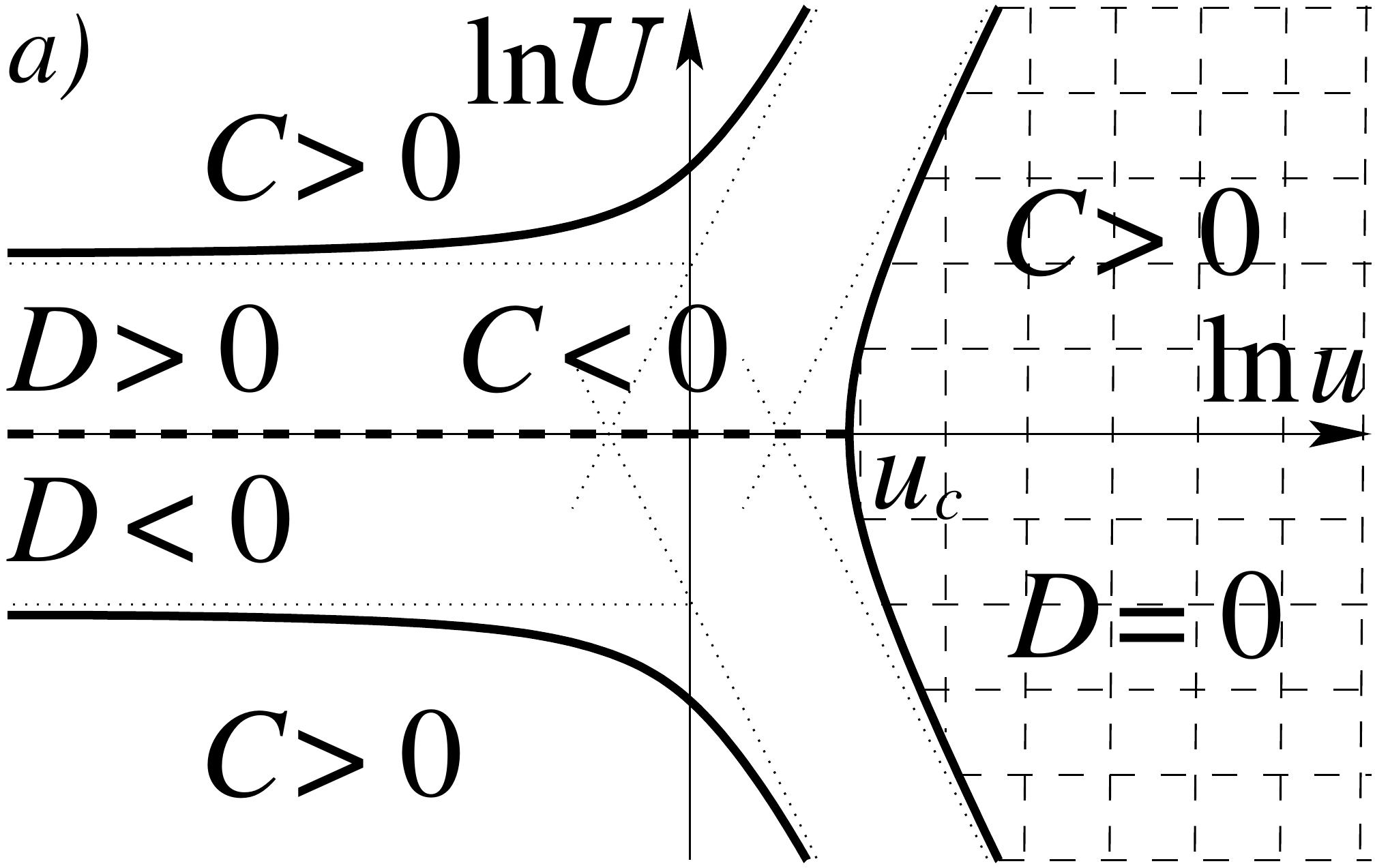}
\qquad\quad
\includegraphics[height=4cm]{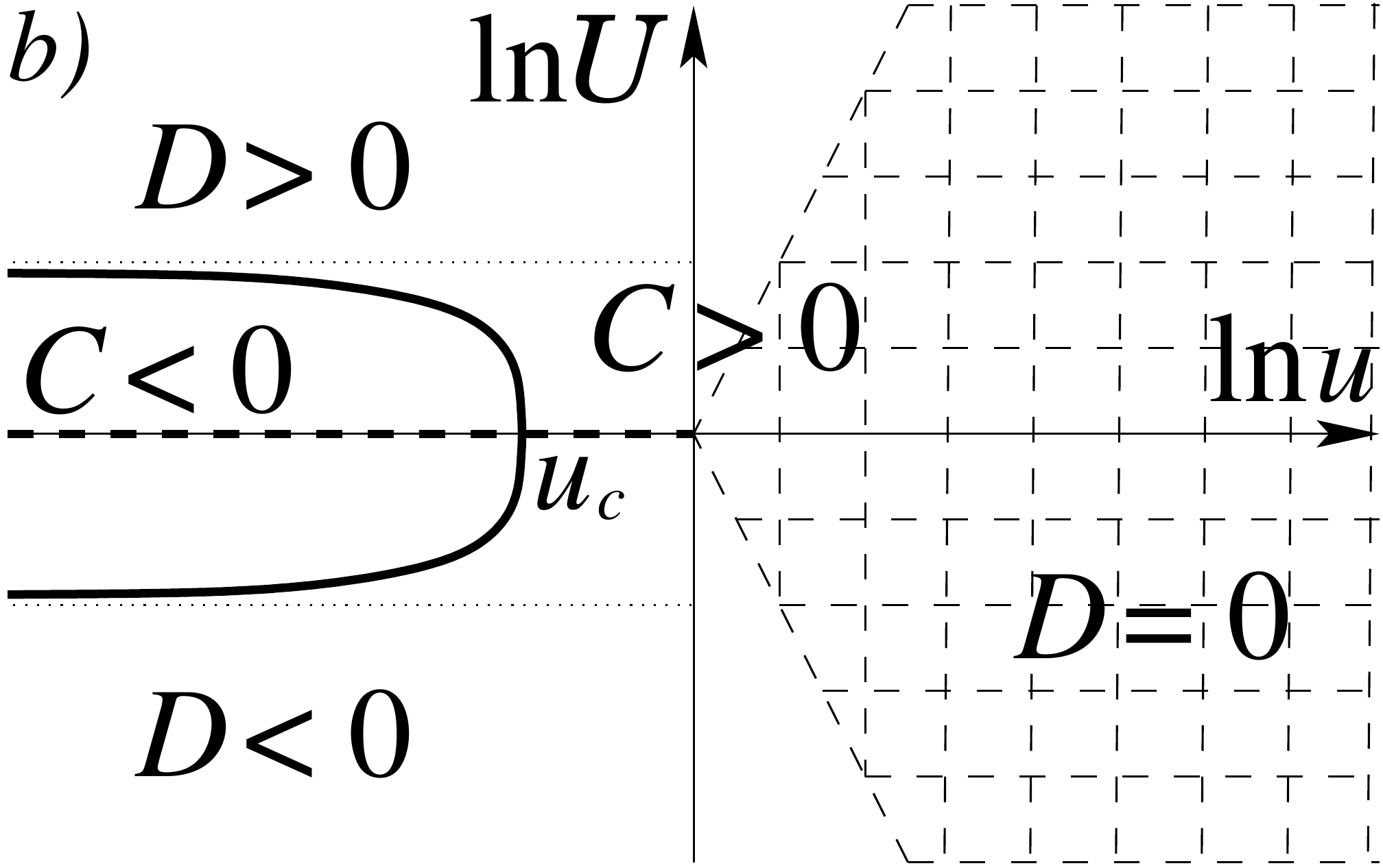}
\caption{\footnotesize \em For a pressure of the form (\ref{case1'}), the curves defined by $C(u,\U)=0$ (straight lines) are represented. The locus $D=0$ is composed of the axis $\U=1$, and the sectors $III$ and $IV$, (see Fig. \ref{sectors}). When $r<r_{c4}$ (Fig. a) or $r_{c4}\leq r<r'_{c3}$ (Fig. b), there is an unstable cosmological solution $(u_c,\U_c=1)$.}
\vspace{-.5cm}
\label{case_1'}
\end{center}
\end{figure}

%%%%%%%%%%%%%
\noindent
$\bullet$ We can treat in a similar way type II {\em Case 1'} models with arbitrary $Q_4+Q_5$. The pressure is of the form
\be
\label{case1'}
p(u,\U)= n_{100}[p_{100}+r p_{111}] \, ,
\ee
where $-1\leq r\leq 1$. This class of models belongs to a segment in the interior of the tetrahedron, $\xi_2=\xi_2={1\over 2}-\xi_1$, $0\leq \xi_1\leq {1\over 2}$. 
The symmetry $\U\to \U^{-1}$ implies $D(u,\U)=0$ along the axis $\U\equiv 1$. Also, $p(u,\U)$ is constant in sectors $III$ and $IV$ (and their common edge $(4)$), implying that $D(u,\U)$ is vanishing. For $r<0$, there is no other solutions to $D=0$.  When $r$ varies, the set of solutions to $C(u,\U)=0$ is divided into 3 classes characterized by
\be
\label{rc'3}
r_{c4}\simeq -0.77 \; , ~~~r'_{c3}=-{f^{ee}_{5/2}(1)\over f^{oo}_{5/2}(1)}\simeq -0.215 \, .
\ee
- For $r<r_{c4}$, the right boundary of the locus $C\le 0$ is asymptotic to the edges $(3)$ and $(5)$, where $D$ is not vanishing yet, (see Fig. \ref{case_1'}{\em a}). The only solution to $D=C=0$ arises at $\U=1$, but the corresponding cosmological evolution is unstable under small fluctuations of $\U$. \\
- For $r_{c4}<r<r'_{c3}$, a cosmological solution with constant $u$ and $\U$ exists, (see Fig. \ref{case_1'}{\em b}), but is again unstable under small fluctuations of $\U$. \\
- For $r'_{c3}<r$,  one has $C(u,\U)>0$ everywhere: There is no cosmological solution with constant complex structures. 
%

\begin{figure}[h!]
\begin{center}
\vspace{.3cm}
\includegraphics[height=4cm]{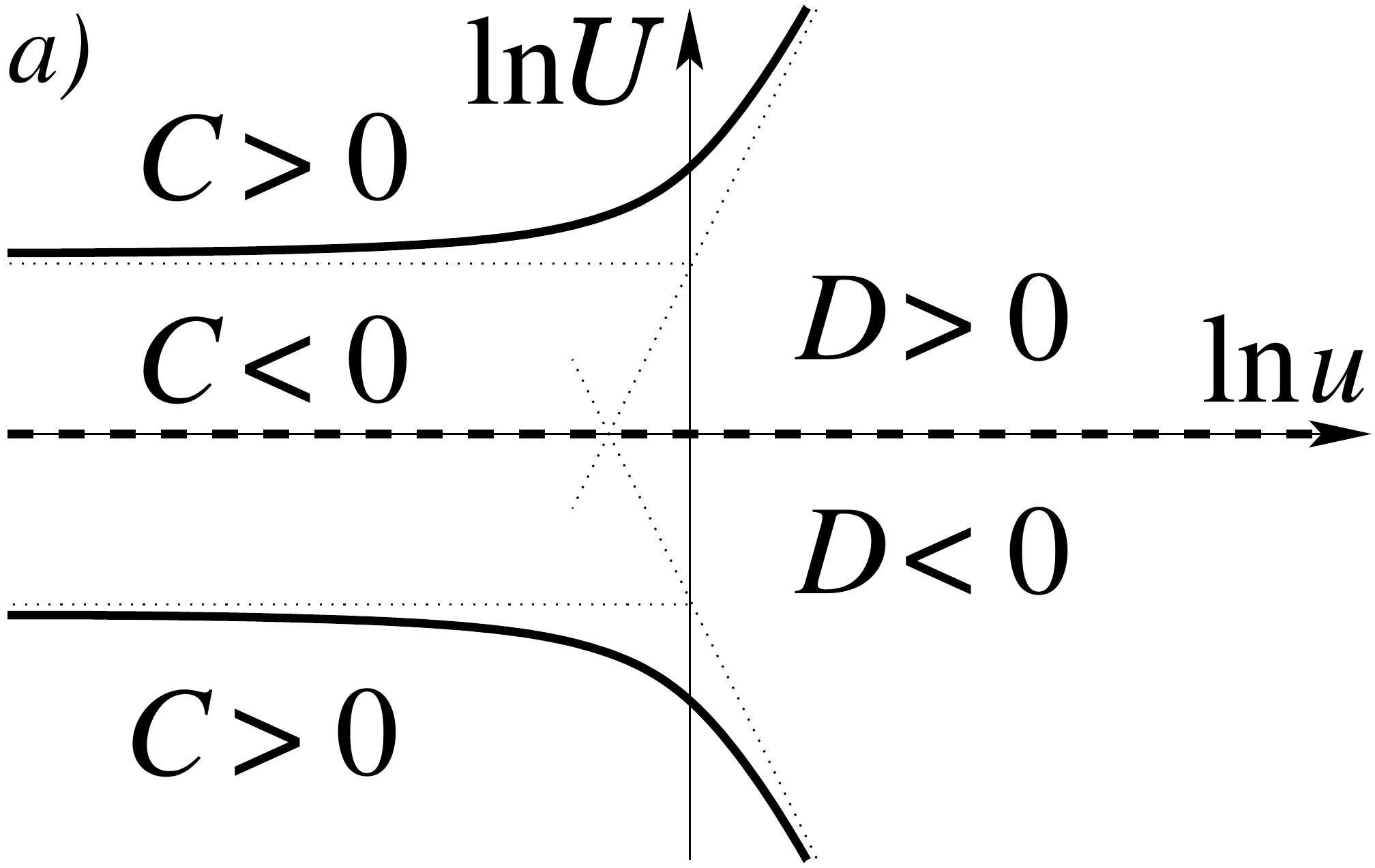}
\qquad\quad
\includegraphics[height=4cm]{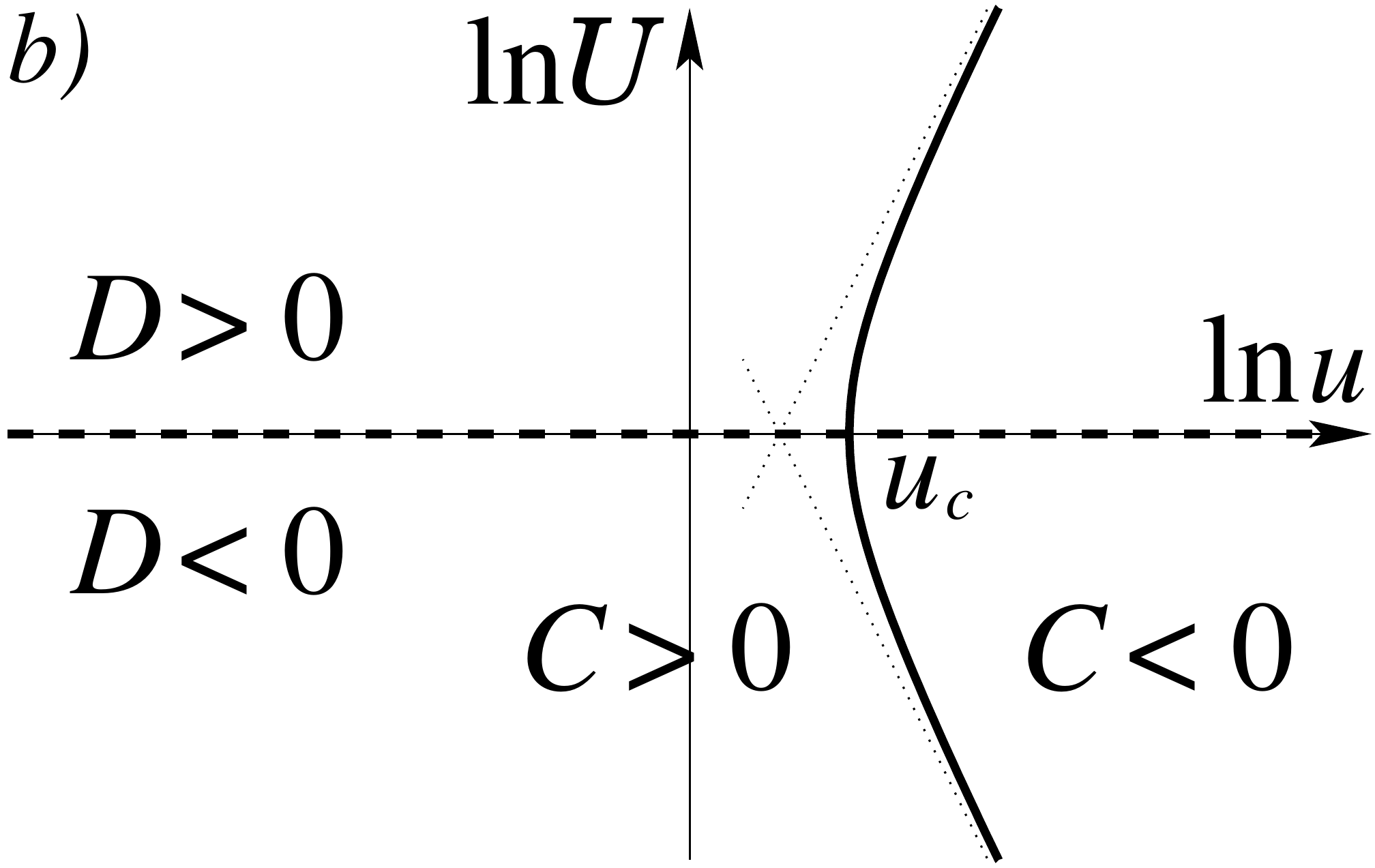}
\caption{\footnotesize \em For a pressure of the form (\ref{case1}), the curves defined by $C(u,\U)=0$ (straight lines) and $D(u,\U)=0$ (dashed lines) are represented. When $r<r'_{c3}$ (Fig. a), there is no cosmological solution with constant complex structures.  When $r'_{c3}\le r<0$ (Fig.  b), there is an unstable cosmological solution $(u_c,\U_c=1)$.}
\vspace{-.5cm}
\label{case_1}
\end{center}
\end{figure}

%%%%%%%%%%%%%
\noindent
$\bullet$ The last  class of systems we analyze corresponds to  type II {\em Case 1} models with arbitrary $Q_4+\bar Q_5$. The pressure is 
 \be
\label{case1}
p(u,\U)= n_{100}[p_{100}+r p_{011}] \, ,
\ee
where $-1\leq r\leq 1$. In contrast to the previous cases, these models are not represented by points in the tetrahedron. For $r<0$, the only solutions to $D(u,\U)=0$ are along the axis $\U=1$. The set of solutions to $C(u,\U)=0$ is characterized by the point $r'_{c3}$ of Eq. (\ref{rc'3}) and $r_{c1}=0$. \\
- For $r<r'_{c3}$, the axis $\U=1$ is entirely inside the region $C(u,\U)<0$ that has two distinct boundaries, (see Fig. \ref{case_1}{\em a}), and so there is no cosmological solution with constant $u$ and $\U$.\\
- For $r'_{c3}\le r<0$, the region $C(u,\U)<0$ has now a connected boundary, (see Fig. \ref{case_1}{\em b}). The latter crosses the axis $\U=1$, so that a cosmological solution $(u_c,\U_c=1)$ exists. It is however unstable under small fluctuations of $\U$.\\
- For $0<r$,  one has $C(u,\U)>0$ everywhere: There is no cosmological solution with constant $u$ and $\U$. 
%

%%%%%%%%%%%%%%%%%%%%%%%%%%%%%%%%%%%%%
%%%%%%%%%%%%%%%%%%%%%%%%%%%%%%%%%%%%%

\subsection{Non-thermal vs thermal cosmologies}
\label{compa}

\noindent
The cosmological evolutions found in the non-thermal and thermal cases,
 Eqs  (\ref{Hubble16}) and (\ref{Hubble4}), have drastically different properties we would like to comment on.

\noindent
First, the contributions of the kinetic energy of the field $\phi_-$ to the ``effective"  Friedmann-Hubble equation are of {\it opposite sign}. This phenomenon is due to the different compatibility relations $\dot   \phi=-4H$ for the non-thermal cosmology and $\dot   \phi=-H$ for the thermal one, once they are inserted in the respective Friedmann-Hubble 
Eqs  (\ref{Hubble}) and (\ref{HubbleT}). 
This is also the reason why these equations have $1/a^{16}$ and $1/a^{4}$ monomial contributions, respectively.
The intermediate cosmological region we are considering is described by the thermal cosmologies.
At late times however, the temperature, being proportional to the inverse of the scale factor, tends to zero,
 and a paradox seems to arise 
as the cosmological evolution is never described by the non-thermal solutions. 

\noindent To better understand this point, we rewrite the pressure and energy density as follows, (see the relations (\ref{defM}) and (\ref{rp})):
\be
P=-M^4\, v_{\mbox{\tiny th}}\, , \qquad \rho=M^4\, (v_{\mbox{\tiny th}}+u\partial_u v_{\mbox{\tiny th}})\qquad \mbox{where}\qquad v_{\mbox{\tiny th}}(u,\U)=-{p(u,\U)\over u^4}\, .
\ee
 When the pressure is of the general form (\ref{pgeneral}), one can use the expansion valid for $u\gg 1$ ($\U$ fixed), Eqs  (\ref{p4}) and (\ref{k}), to decompose $v_{\mbox{\tiny th}}$ into two pieces:
\be
\label{vdecompo}
v_{\mbox{\tiny th}}(u,\U)=-\left\{n_{010}k_{3}^{oe}(\U^{-1})+n_{001}k_{3}^{oe}(\U)+n_{011}k_{3}^{oo}(\U)\right\}+\hat v_{\mbox{\tiny th}}(u,\U)\, ,
\ee
where 
\be
\hat v_{\mbox{\tiny th}}(u,\U)=-{1\over u^4}\times \left\{(n_{100}+n_{111})S_4^o+(n_{010}+n_{001}+n_{011})S_4^e\right\}+\cdots\, ,
\ee
up to exponentially suppressed terms as $u\to +\infty$. 
Clearly, $\hat v_{\mbox{\tiny th}}(u,\U)\to 0$ in this limit. Since $n_{\tg_0\tg_4\tg_5}=n_{\tg_4\tg_5}$, 
the $u$-independent terms of the r.h.s. of Eq. (\ref{vdecompo}) are equal to the zero temperature effective potential 
$v(\U)$ found for the heterotic or type II cases, Eqs  (\ref{vhet}) and (\ref{vII}).  
Thus, $P$ and $\rho$ are only converging to their $T= 0$ counterparts when $u\to +\infty$. 
However, the thermal cosmologies we have considered have stabilized $u\equiv u_c$, 
implying that the finite temperature corrections $\hat v_{\mbox{\tiny th}}(u,\U)$, even if the temperature is small, are never negligible. 
In fact, the condition $u= M(t)/T(t) \to+\infty$ for the thermal system to be correctly approximated by the non-thermal one implies  that  the thermal corrections should be screened by radiative corrections, and not that they would necessarily vanish. 
Explicit cosmological evolutions with $u\to+\infty$ are analyzed in \cite{BKP}. They describe Big Crunch cosmologies, where the temperature 
is proportional to $1/a$. So, $T$ is large in absolute value (but still negligible compared to $M$).

\noindent Finally, we note that in the present work, each time a stable cosmological solution at finite temperature is found, its counterpart at $T= 0$ is unstable, and vice versa. As explained in \cite{BKP}, this is a consequence of the fact that for arbitrary initial conditions, the cosmological evolutions are always converging to an attractor.  Depending on the model, the latter can be the radiation dominated universes studied in \cite{ckpt} and the present paper, or Big Crunch cosmologies where the thermal effects are screened by radiative corrections. 

%%%%%%%%%%%%%%%%%%%%%%%%%%%%%%%%%%%%%%%
%%%%%%%%%%%%%%%%%%%%%%%%%%%%%%%%%%%%%%%

\section{Conclusions}

The main result of this work is to show the existence of the critical stringy cosmologies of \cite{ckpt}, even in the cases where more than one modulus participate in the supersymmetry breaking mechanism. They correspond to a radiation dominated era with constant complex structures.
A thorough analysis was done for several string models with $N_4\ge 2$ initial supersymmetry. Depending on the pattern of supersymmetry breaking, the critical values of the thermal effective potential for the complex structure moduli are either minima or  maxima. Run away behaviors that bring the system to higher dimensions can also occur. 
When the radiation era is stable, we explicitly show that even if the temperature tends to zero at late times, thermal corrections to the dynamics are never negligible.

\noindent
Our approach was to separate the cosmological evolution in distinct regions, according to the value of the temperature. The stringy non-geometrical region, where the temperature is of order the Hagedorn temperature, is much harder to understand. Stringy phenomena occur where conventional field theoretic notions concerning the geometry and topology are breaking down. Some interesting proposals to understand this early time region have been  put forward in \cite{akpt,massivesusy}.
The intermediate region, just after the Hagedorn era is shown to be under control. The free energy is free of any infrared and ultraviolet ambiguities, allowing us to follow the backreaction on the geometry and determine the time evolution of several moduli fields. We show that the only relevant moduli are the ones participating in the supersymmetry breaking mechanism. The others are either frozen by receiving a soft-breaking mass, or they remain flat directions with exponentially suppressed contribution to the free energy.

\noindent
A very interesting result is that the critical cosmological solutions are governed by a higher dimensional thermal state equation, $\rho=(d-1)\,P$, or in the four-dimensional effective description, this equation becomes $\rho=(3+n)\,P$ where $n$ is the number of moduli participating in the breaking of supersymmetry. In \cite{ckpt,ak}, we analyzed the case $n=1$. Here, we generalize the result to the $n=2$ case. The overall scale evolves in time such that the state equation is valid, while complex structure-like ratios of scales are frozen. 

\noindent
Although our work covers adequately and unambiguously the intermediate cosmological region, it is incomplete to describe the very early Hagedorn era, where non-geometric stringy phenomena are important. It is incomplete as well to describe relatively late time phenomena, like the radiative breaking of the electroweak gauge symmetry and QCD confinement, where non-perturbative transmutation scales, like for instance $\Lambda_{QCD}$, are relevant. The extension to the late time era requires to consider at least models with $N_4=1$ initial supersymmetry. Progress concerning the very early era can be made provided that we understand better the resolution of the Hagedorn instabilities and the stringy non-geometrical structure of the early universe.        

%%%%%%%%%%%%%%%%%%%%%%%%%%%%%%%%%%%%
\section*{Acknowledgements}

We are grateful to C. Angelantonj, C. Bachas, J. Estes, J. Troost and especially F. Bourliot
for useful discussions.
H.P. and N.T. thank the Ecole Normale Sup\'erieure and C.K. and H.P. the University of Cyprus for hospitality.\\
\noindent The work of C.K. and H.P. is partially supported by the EU contract MRTN-CT-2004-005104
and the ANR (CNRS-USAR) contract  05-BLAN-0079-01 (01/12/05). N.T. and C.K. are supported by the EU
contract MRTN-CT-2004-512194. H.P. is also supported by the EU contracts MRTN-CT-2004-503369 and
MEXT-CT-2003-509661, INTAS grant 03-51-6346, and CNRS PICS 2530, 3059 and 3747, while N.T. is also supported
by an INTERREG IIIA Crete/Cyprus program.

%%%%%%%%%%%%%%%%%%%%%%%%%%%%%%%%%%%%

\section*{Appendix A}
%\begin{center}
%\Large\bf Appendix A}
%\end{center}
\renewcommand{\theequation}{A.\arabic{equation}}
\renewcommand{\thesection}{A.}
\setcounter{equation}{0}

\noindent 
We list in this Appendix the large and small complex structure expansions of an arbitrary linear combination of functions $p_{\tg_0\tg_4\tg_5}$ defined in Eq. (\ref{pggg}),
\be
\label{pgeneral}
p(u,\U)=n_{100}( p_{100}+r_{010}p_{010}+r_{001}p_{001}+r_{111}p_{111}+r_{011}p_{011})\,.
\ee
For fixed $\omega$ and $\lambda$, we determine the power expansion along the lines $\ln\U=\omega\ln u+\ln \lambda$ when $\ln u$ and/or $\ln \U$ are large, by making an extensive use of the approximations
\be
\begin{array}{l}
\displaystyle\sum_m {1\over \left((2m+1)^2 x+ a\right)^3}={3\pi\over 16}\, {1\over a^{5/2}}\, {1\over \sqrt{x}}+\cdots \, ,\\
\displaystyle\sum_m {1\over \left((2m)^2 x+ a\right)^3}={3\pi\over 16}\, {1\over a^{5/2}}\, {1\over \sqrt{x}}+\cdots \, , 
\end{array}
\ee 
where $a>0$, $x\to 0_+$ and the dots stand for $\O\left({e^{-\pi \sqrt{a/ x}}\left/ x^{3/2}\right.}\right)$ terms. Similarly, we use    
\be
\begin{array}{l}
\displaystyle\sum_m {1\over \left((2m+1)^2 x+ a\right)^{5/2}}={2\over 3}\, {1\over a^{2}}\, {1\over \sqrt{x}}+\cdots \, ,\\
\displaystyle\sum_m {1\over \left((2m)^2 x+ a\right)^{5/2}}={2\over 3}\, {1\over a^{2}}\, {1\over \sqrt{x}}+\cdots\, ,
\end{array}
\ee 
where again we neglect exponentially suppressed terms. The $(\ln u,\ln \U)$-plane can then be divided in 6 sectors, $I$, $II$, $\dots$, $VI$, where the expansions are independent of $\omega$ (and $\lambda$). The boundaries of these sectors are the lines whose slopes are $\omega=-2,0$ or 2, (see Fig. \ref{sectors}). In each sector, we find,
\be
\!\!\!\!\!\begin{array}{lll}
\displaystyle {p^{I}\over n_{100}}&=&u^{-2}~S_6^o \\ 
&&+ u^3\,\U^{-5/2}(S_5^e+r_{001}S_5^o)\\
&&+u^3\, \U^{3/2}((1+r_{010}+r_{001})S_4^e+(r_{111}+r_{011})S_4^o)\,, 
\end{array}
\ee
\be
\!\!\!\!\begin{array}{lll}
\displaystyle {p^{II}\over n_{100}}&=&u^4\,\U^{-3}~r_{001}S_6^o\\
&&+u^{-1}\U^{-1/2}~(S_5^o+r_{001}S_5^e)\\
&&+u^3\,\U^{3/2}~((1+r_{001})S_4^e+(r_{010}+r_{111}+r_{011})S_4^o)\,,
\end{array}
\ee
\be
\!\!\!\!\!\!\!\!\!\!\!\!\!\!\!\!\!\!\!\!\!\!\!\!\!\!\begin{array}{lll}
\displaystyle {p^{III}\over n_{100}}&=&u^4\,\U^{-3}~r_{001}S_6^o\\
&&+u^{4}\,\U^{2}~((r_{010}+r_{011})S_5^o+r_{001}S_5^e)\\
&&+(1+r_{111})S_4^o+(r_{010}+r_{001}+r_{011})S_4^e\,,
\end{array}
\ee
\be
\!\!\!\!\!\!\!\!\!\!\!\!\!\!\!\!\!\!\!\!\!\!\!\!\!\!\begin{array}{lll}
\displaystyle {p^{IV}\over n_{100}}&=&u^4\,\U^{3}~r_{010}S_6^o\\
&&+u^{4}\,\U^{-2}~((r_{001}+r_{011})S_5^o+r_{010}S_5^e)\\
&&+(1+r_{111})S_4^o+(r_{010}+r_{001}+r_{011})S_4^e\,,
\end{array}
\ee
\be
\!\begin{array}{lll}
\displaystyle {p^{V}\over n_{100}}&=&u^4\,\U^{3}~r_{010}S_6^o\\
&&+u^{-1}\U^{1/2}~(S_5^o+r_{010}S_5^e)\\
&&+u^3\,\U^{-3/2}~((1+r_{010})S_4^e+(r_{001}+r_{111}+r_{011})S_4^o)\,,
\end{array}
\ee
\be
\begin{array}{lll}
\displaystyle {p^{VI}\over n_{100}}&=&u^{-2}~S_6^o \\ 
&&+ u^3\,\U^{5/2}~(S_5^e+r_{010}S_5^o)\\
&&+u^3\,\U^{-3/2}~((1+r_{010}+r_{001})S_4^e+(r_{111}+r_{011})S_4^o)\,,
\end{array}
\ee
where we have defined 
\be
\begin{array}{lll}
\displaystyle S_6^o={2\over \pi^3}\sum_m {1\over (2m+1)^6}={\pi^3\over 240}\,,&\\
\displaystyle S_5^o={2\over \pi^3}\; {3\pi\over 16}\sum_m {1\over |2m+1|^5}={93\over 128}\; {\zeta(5)\over \pi^2}&,& \displaystyle S_5^e={2\over \pi^3}\; {3\pi\over 16}\sum_{m\neq 0} {1\over |2m|^5}={3\over 128}\; {\zeta(5)\over \pi^2}\,,\\
\displaystyle S_4^o={2\over \pi^3}\; {3\pi\over 16}\; {2\over 3}\sum_m {1\over (2m+1)^4}={\pi^2\over 192}&,& \displaystyle S_4^e={2\over \pi^3}\; {3\pi\over 16}\; {2\over 3}\sum_{m\neq 0} {1\over (2m)^4}={\pi^2\over 2880}\,.
\end{array}
\ee
\noindent 
The previous sectors are separated by edges, $(1),\dots,(6)$, in the neighborhood of which some terms we neglected in the interior of the adjacent sectors are not exponentially suppressed anymore. Along these edges,  one has $\U=\lambda u^{\pm 2,0}$, where $\lambda \simeq 1$, and
$$
\!\!\!\!\!\!\!\!\!\!\!\!\!\!\!\begin{array}{lll}
\displaystyle {p^{(1)}\over n_{100}}&=& u^{-2}~S_6^o\\
&& +u^3~(f_{5/2}^{ee}(\lambda)+r_{010}f_{5/2}^{oe}(\lambda^{-1})+r_{001}f_{5/2}^{oe}(\lambda)+(r_{111}+r_{011})f_{5/2}^{oo}(\lambda))\;,~\lambda=\U,
\end{array}
$$
$$
\!\!\!\!\!\!\!\!\!\!\!\!\!\!\!\!\!\!\!\!\!\!\!\!\!\!\!\!\!\!\!\!\!\!\!\!\!\!\!\!\!\!\!\!\!\!\!\!\!\!\!\!\!\!\!\!\!\!\!\!\!\!\!\!\!\!\!\!\!\!\!\!\!\!\!\!\!\!\!\!\!\!\!\!\!\!\!\!\!\begin{array}{lll}
\displaystyle {p^{(2)}\over n_{100}}&=& u^{3}\,\U^{3/2}~((1+r_{001})S_4^e+(r_{010}+r_{111}+r_{011})S_4^o)\\
&& +u^{-2}~(g_{3}^{eo}(\lambda)+r_{001}g_{3}^{oe}(\lambda))\;,~\lambda=\U u^{-2},
\end{array}
$$
$$
\!\!\!\!\!\!\!\!\!\!\!\!\!\!\!\!\!\!\!\!\!\!\!\!\!\!\begin{array}{lll}
\displaystyle {p^{(3)}\over n_{100}}&=& u^{4}\,\U^{-3}~r_{001}S_6^o\\
&& +h_{5/2}^{oe}(\lambda)+(r_{010}+r_{011})h_{5/2}^{eo}(\lambda)+r_{001}h_{5/2}^{ee}(\lambda)+r_{111}h_{5/2}^{oo}(\lambda)\;,~\lambda=\U u^2,
\end{array}
$$
$$
\!\!\!\!\!\!\!\!\!\!\!\!\!\!\!\!\!\!\!\!\!\!\!\!\!\!\!\!\!\!\!\!\!\!\!\!\!\!\!\!\!\!\!\!\!\!\!\!\!\!\!\!\!\!\!\!\!\!\!\!\!\!\!\!\!\!\!\begin{array}{lll}
\displaystyle {p^{(4)}\over n_{100}}&=& (1+r_{111})S_4^o+(r_{010}+r_{001}+r_{011})S_4^e\\
&& +u^{4}~(r_{010}k_{3}^{oe}(\lambda^{-1})+r_{001}k_{3}^{oe}(\lambda)+r_{011}k_{3}^{oo}(\lambda))\;,~\lambda=\U,
\end{array}
$$
$$
\begin{array}{lll}
\displaystyle {p^{(5)}\over n_{100}}&=& u^{4}\U^{3}~r_{001}S_6^o\\
&& +h_{5/2}^{oe}(\lambda^{-1})+r_{010}h_{5/2}^{ee}(\lambda^{-1})+(r_{001}+r_{011})h_{5/2}^{eo}(\lambda^{-1})+r_{111}h_{5/2}^{oo}(\lambda^{-1})\;,~\lambda=\U u^{-2},
\end{array}
$$
$$
\!\!\!\!\!\!\!\!\!\!\!\!\!\!\!\!\!\!\!\!\!\!\!\!\!\!\!\!\!\!\!\!\!\!\!\!\!\!\!\!\!\!\!\!\!\!\!\!\!\!\!\!\!\!\!\!\!\!\!\!\!\!\!\!\!\!\!\!\!\!\!\!\!\!\!\!\!\!\!\!\!\!\!\!\!\!\begin{array}{lll}
\displaystyle {p^{(6)}\over n_{100}}&=& u^{3}\U^{-3/2}~((1+r_{010})S_4^e+(r_{001}+r_{111}+r_{011})S_4^o)\\
&& +u^{-2}~(g_{3}^{eo}(\lambda^{-1})+r_{010}g_{3}^{oe}(\lambda^{-1}))\;,~\lambda=\U u^{2},
\end{array}
$$
\be\label{p4}\ee
where we have introduced the functions
\be
\label{ffunc}
\begin{array}{l}
\displaystyle f_{5/2}^{ee}(\lambda)={2\over \pi^3}\, {3\pi\over 16}\sum_{(m,n)\neq(0,0)}{1\over ((2m)^2\lambda +(2n)^2\lambda^{-1})^{5/2}}\, ,\\
\displaystyle f_{5/2}^{oe}(\lambda)={2\over \pi^3}\, {3\pi\over 16}\sum_{m,n}{1\over ((2m+1)^2\lambda +(2n)^2\lambda^{-1})^{5/2}}\, ,\\
\displaystyle f_{5/2}^{oo}(\lambda)={2\over \pi^3}\, {3\pi\over 16}\sum_{m,n}{1\over ((2m+1)^2\lambda +(2n+1)^2\lambda^{-1})^{5/2}}\, ,
\end{array}
\ee
\be
\begin{array}{ll}
\displaystyle g_{3}^{eo}(\lambda)={2\over \pi^3}\sum_{m,n}{1\over ((2m)^2\lambda +(2n+1)^2)^{3}}\, ,&
\displaystyle g_{3}^{oe}(\lambda)={2\over \pi^3}\sum_{m,n}{1\over ((2m+1)^2\lambda +(2n)^2)^{3}}\, ,\end{array}
\ee
\be
\begin{array}{l}
\displaystyle  h_{5/2}^{oe}(\lambda)={2\over \pi^3}\, {3\pi\over 16}\, \lambda^2\sum_{m,n}{1\over ((2m+1)^2\lambda +(2n)^2)^{5/2}}\, ,\\\displaystyle  h_{5/2}^{eo}(\lambda)={2\over \pi^3}\, {3\pi\over 16}\, \lambda^2\sum_{m,n}{1\over ((2m)^2\lambda +(2n+1)^2)^{5/2}}\, ,\\
\displaystyle  h_{5/2}^{ee}(\lambda)={2\over \pi^3}\, {3\pi\over 16}\, \lambda^2\sum_{(m,n)\neq(0,0)}{1\over ((2m)^2\lambda +(2n)^2)^{5/2}}\, ,\\ \displaystyle  h_{5/2}^{oo}(\lambda)={2\over \pi^3}\, {3\pi\over 16}\, \lambda^2\sum_{m,n}{1\over ((2m+1)^2\lambda +(2n+1)^2)^{5/2}}\, ,
\end{array}
\ee
\be
\label{k}
\begin{array}{ll}
\displaystyle k_{3}^{oe}(\lambda)={2\over \pi^3}\sum_{m,n}{1\over ((2m+1)^2\lambda +(2n)^2\lambda^{-1})^{3}}\, ,&
\displaystyle k_{3}^{oo}(\lambda)={2\over \pi^3}\sum_{m,n}{1\over ((2m+1)^2\lambda +(2n+1)^2\lambda^{-1})^{3}}\, .\end{array}
\ee

%%%%%%%%%%%%%%%%%%%%%%%%%%%%%%%%%%%%%%%%%
%%%%%%%%%%%%%%%%%%%%%%%%%%%%%%%%%%%%%%%%%

\end{document}